\documentclass[journal]{IEEEtran}

\usepackage{graphicx}
\usepackage{amsmath}
\usepackage{url}
\usepackage{dsfont}
\usepackage{mathtools}

\DeclareMathOperator*{\argmin}{arg\,min}

\usepackage{color}

\begin{document}

\title{Multitask Learning for Fundamental Frequency Estimation in Music}

\author{Rachel M. Bittner$^1$,
        Brian McFee$^1$,
		Juan P. Bello$^1$\\
		$^1$ Music and Audio Research Lab, New York University, USA}


\maketitle

\begin{abstract}
Fundamental frequency ($f_0$) estimation from polyphonic music includes the tasks of multiple-f0, melody, vocal, and bass line estimation.
Historically these problems have been approached separately, and only recently, using learning-based approaches.
We present a multitask deep learning architecture that jointly estimates outputs for various tasks including multiple-f0, melody, vocal and bass line estimation, and is trained using a large, semi-automatically annotated dataset.
We show that the multitask model outperforms its single-task counterparts, and explore the effect of various design decisions in our approach, and show that it performs better or at least competitively when compared against strong baseline methods.
\end{abstract}


\IEEEpeerreviewmaketitle

\section{Introduction}
The fundamental frequency ($f_0$) or ``pitch'' of a musical sound is a perceptual property that allows us to order sounds from low to high.
In music, the $f_0$ information varies over time and is a property of the parts played by various voices and musical instruments.
The process of extracting this information from musical audio is known as \emph{$f_0$ estimation}, and is one of the foundational problems in the field of Music Information Retrieval (MIR).
From a musical perspective, pitch and rhythm are the basis of virtually all music across cultures, and this is reflected in nearly all graphical representations of music.
From an evolutionary perspective, pitch is central to human communication and survival.
In speech, consider the difference between the way the sentences ``You have a car.'' and ``You have a car?'' would be spoken --- the primary difference is the variation in pitch over time.
In non-spoken verbal communication, pitch cues are used to convey information such as emotions (e.g. affirmation, displeasure) or the presence of danger (e.g. screaming).
From a psycho-physical perspective, the human ear contains the cochlea --- an organ specifically devoted to decoding pitch information.

$f_0$ estimation has a diverse set of applications. Within MIR, it is used as vital side information for tasks such as informed source separation and analysis/re-synthesis~\cite{kawahara2016using}. In music production, $f_0$ estimation is at the core of commercial applications such as Melodyne or AutoTune which perform pitch correction or modification. In music theory, performance and education, a live recordings or recorded work that does not exist in written form can be transcribed for further analysis or practice~\cite{devaney2008empirical}. $f_0$ estimation is used in musicology for large-scale analysis of different musical styles~\cite{panteli2017towards}, and it has been used for music similarity and retrieval tasks, such as “query by humming” (i.e. search for a song by humming the melody), and version/cover song detection~\cite{salamonVersionIDQBH-MMIR13}.
$f_0$ curves, specifically melody curves, have also been used as a feature for genre identification~\cite{salamon2012musical}.

$f_0$ estimation is a difficult task due to a number of factors including interference from percussive sources, varying patterns of harmonics, and high degrees of polyphony resulting in overlapping harmonics.
These problems are made even more prevalent in music with dense mixtures such as in heavily produced pop music with substantial mastering effects.
This challenges have begun to be addressed recently with the use of data-driven approaches.
However, data-driven approaches require large amounts of labeled data --- in this case, musical audio with time aligned $f_0$ labels.
Real music with labeled $f_0$ information is in short supply, as it requires considerable human labeling effort~\cite{bittner2014medleydb,Mauch:2014gm}.
As a result, models for each $f_0$ estimation task rely on small, task-specific datasets both for training and testing, and many of these datasets are regrettably homogeneous.

In this work, we propose the use of multitask learning as a way to address the scarcity of $f_0$ data by exploiting the commonalities between the various $f_0$ estimation tasks.
More specifically, we propose a novel multitask neural network architecture and various innovative strategies for data annotation and synthesis that enable the training of this model.
We demonstrate that the proposed multitask architecture performs better than the equivalent single task architectures; that the more tasks used in the multitask architecture, the better the performance, even when some tasks are trained with purely synthesized data; and that, given a good match between the training and testing data, the multitask model performs as well or better than existing state-of-the-art task-specific models.

The remainder of this paper is organized as follows.
In Section~\ref{sec:litreview}, we give an overview of the current state of the art for each $f_0$ estimation task, as well as the current state of multitask learning.
In Section~\ref{sec:approach} we describe the details of the proposed multitask model.
In Section~\ref{sec:training-data}, we explain how we merge existing datasets and leverage synthesized audio to create a large enough training set.
In Section~\ref{sec:experiments-setup} we describe the evaluation data, metrics, and the set of experiments we perform.
In Section~\ref{sec:experiments}, we compare our proposed multitask architecture with equivalent single task architectures, examine the effect of several of our design choices, and compare our approach with strong baselines.
We conclude in Section~\ref{sec:conclusions} with limitations and future directions.

\section{Related Work}\label{sec:litreview}
$f_0$ estimation has been studied in various forms for almost 40 years, with most earlier work focusing on estimating the pitch of a single, monophonic source, and later work focusing on estimating $f_0$ information from polyphonic mixtures.
Until recently, the majority of methods for all of these problems were signal processing algorithms built to exploit assumptions about the structure of the input data, for example, that the waveforms are periodic~
\cite{deCheveigne:2002hd}. 
For thorough literature reviews in each of these areas, we refer the reader to review articles on monophonic pitch tracking~\cite{hessPitchSpeech83}, melody estimation~\cite{Salamon:2014fh}, and on multiple-$f_0$ estimation~\cite{benetos2013automatic}.
These methods typically work well for the particular type of musical source the system was designed for, but do not usually generalize well to different types of music~\cite{bittner2014medleydb}.

Historically there has been very little ground truth data available to train and test algorithms.
Recently, more $f_0$ data has become available~\cite{bittner2014medleydb,bosch2016evaluation,chan2015vocal,su2015escaping,emiya2008transcription,pfleiderer2010jazzomat}, sparking the wide-spread adoption of data-driven models for $f_0$ estimation.
These data-driven models have shown promising results and have improved on the state of the art across all areas of $f_0$ estimation.
For monophonic pitch tracking, a convolutional recurrent neural network outperformed the current state of the art expert system pitch tracker~\cite{kim2018crepe}.
For vocal $f_0$ estimation, a Bidirectional LSTM architecture has been proposed~\cite{rigaud2016singing} as well as a multi-scale convolutional architecture with a ``harmonic'' loss function~\cite{kum2016melody}.
Fully connected deep architectures have been used for melody~\cite{balke2017data} and bass extraction~\cite{abesser2017bass} in jazz music.
Fully convolutional architectures have been proposed for melody and multiple-$f_0$ estimation~\cite{bittner2017deep}.
For multiple-$f_0$ tracking, different flavors of recurrent neural networks have been applied to solo piano transcription~\cite{hawthorne2018onsets,sigtia2016end,bock2012polyphonic}.

While each of the previously mentioned models show promising results, they are each built for a single type of $f_0$ estimation, and are trained on the limited data available for the corresponding task.
$f_0$ estimation problems, such as melody, bass and multiple-$f_0$ estimation are inherently related, and data from one task could be used to inform models for another.
Multitask learning~\cite{caruana1998multitask} is a class of machine learning model that uses shared information to predict multiple outputs.
Multitask modeling is beneficial in the context of deep neural networks in particular because it allows different tasks to share latent information.
For a recent overview of the use of multitask learning models we refer the reader to two review articles~\cite{zhang2017multitask,ruder2017multitask}.
Multitask architectures~\cite{caruana1998multitask} have shown improvement in domains including natural language processing~\cite{collobert2008unified}, computer vision~\cite{kendall2017multi}, and speech recognition~\cite{deng2013new}.
They have also been proposed for a handful of music related tasks and shown improvement over single-task models.

Beat and downbeat tracking has seen improvement by jointly estimating both tasks~\cite{bock2016joint}, and even further improvement when additionally transcribing drumming patterns~\cite{vogl2107drum}.
Similarly, chord and key estimation has been shown to improve when estimated jointly~\cite{mauch2010simultaneous}, and the addition of joint downbeat estimations further benefits results~\cite{papadopoulos2011joint}.
For large vocabulary chord recognition, a structured multitask model predicting root, bass, and quality jointly achieved state-of-the-art results and improved over an equivalent single-task, multi-class model~\cite{mcfee2017_structured}.
Multitask learning has not yet been applied to $f_0$ estimation, and crucially, would allow us to combine the limited data from different $f_0$ estimation tasks to address the issue of data scarcity that limits the application of data-driven methods.

\section{Approach}\label{sec:approach}
\begin{figure}
\begin{center}
\includegraphics[width=\columnwidth]{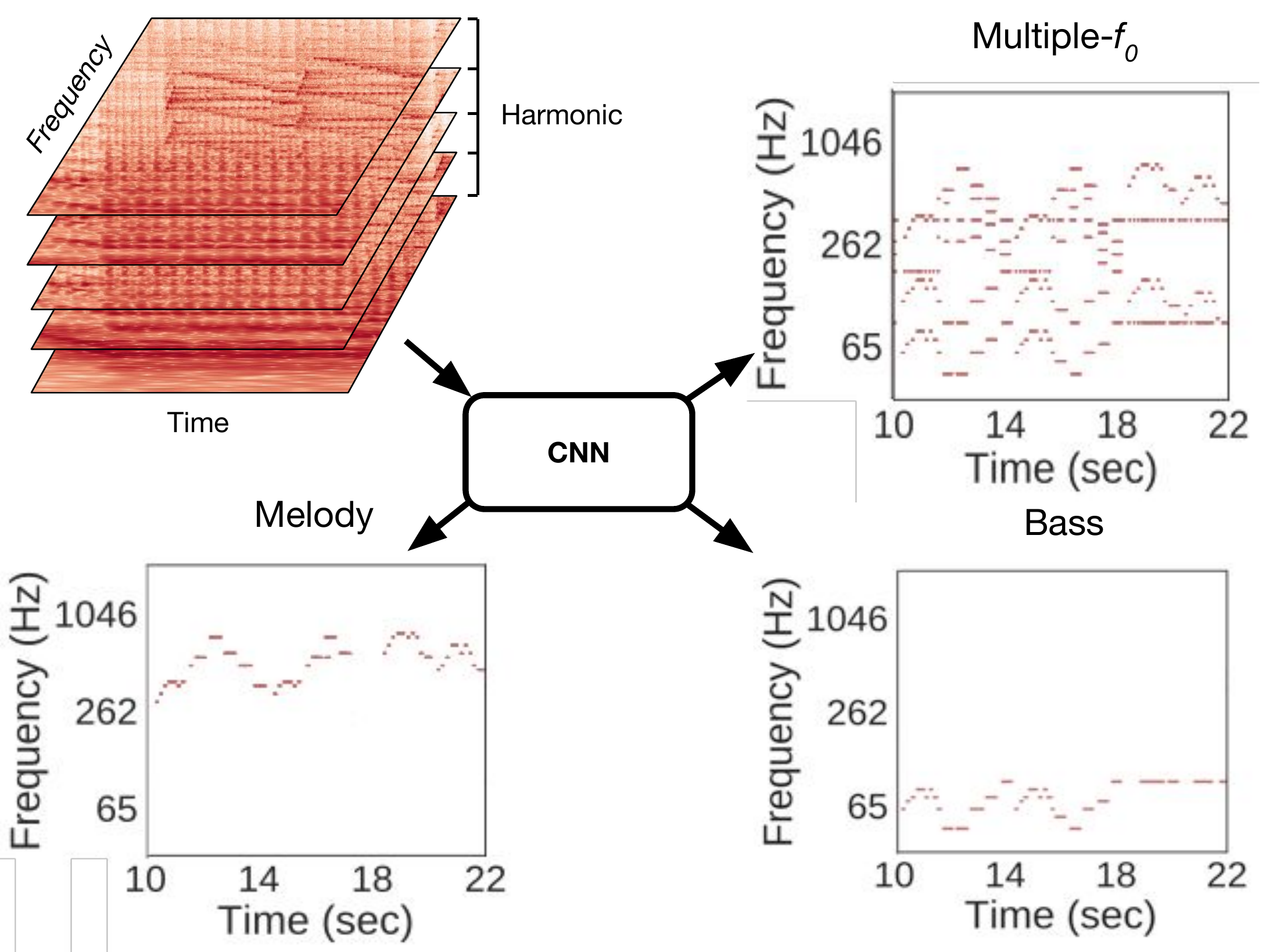}
\caption{Illustration of the proposed system. A CNN is trained to predict ideal salience representations for several $f_0$ estimation tasks from an input HCQT.\label{fig:input-output}}
\end{center}
\end{figure}

In this section, we describe our multitask learning approach to $f_0$ estimation.
Our solution is based on fully convolutional neural network architectures which map a Harmonic Constant-Q Transform (HCQT) input to one or more ideal pitch salience outputs corresponding to specific $f_0$ estimation tasks, as illustrated in Figure~\ref{fig:input-output}.

\subsection{Input Representation}\label{subsec:input-rep}
Pitched musical content often has a harmonic frequency structure, i.e. there is spectral energy at integer multiples of the $f_0$.
To capture this relationship, we use the HCQT~\cite{bittner2017deep} as our input representation.
The HCQT is a 3-dimensional array indexed by harmonic, frequency, and time: $\mathcal{H}[h,t,f]$, measures harmonic $h$ of frequency $f$ at time $t$.
The harmonic $h=1$ refers to the fundamental, and ... $\mathcal{H}[h]$ denotes harmonic $h$ of the ``base'' CQT $\mathcal{H}[1]$.
For any harmonic $h>0$, $\mathcal{H}[h]$ is computed as a standard CQT where the minimum frequency is scaled by the harmonic: $h\cdot f_{\text{min}}$, and the same frequency resolution and number of octaves is shared across all harmonics.
The resulting representation $\mathcal{H}$ is similar to a color image, where the $h$ dimension is the \emph{depth}.
In this work, we compute $\mathcal{H}$ for $h=1,\cdots,5$.

\begin{figure}
\begin{center}
\includegraphics[width=\columnwidth]{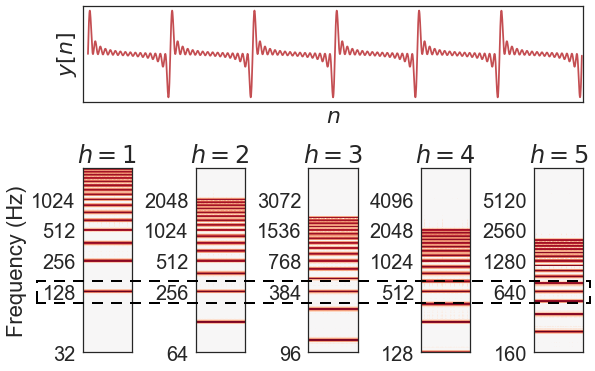}
\caption{(Top) A perfectly harmonic signal $y[n]$ with 16 harmonics and a $f_0$ of 128 Hz. (Bottom) HCQT of $y[n]$ with $\mathcal{H}[1]$ through $\mathcal{H}[5]$ shown. Reading from left to right, the $f_0$ of 128 Hz and its first four harmonics are aligned across the $h$ dimension, outlined by a black dotted line.}\label{fig:hcqt-ex}
\end{center}
\end{figure}

To illustrate this further Figure~\ref{fig:hcqt-ex} (top) shows a perfectly harmonic signal
\begin{equation}
y[n] = \sum_{h=1}^{16} \sin\left(2 \pi \cdot 128 h \cdot \frac{n}{f_s} \right)
\end{equation}
which has an $f_0$ of 128 Hz and 15 harmonics above the fundamental.
Figure~\ref{fig:hcqt-ex} (bottom) shows $\mathcal{H}[1]$ through $\mathcal{H}[5]$.
Note that harmonically related frequencies are aligned in the $h$ dimension -- in particular the $f_0$ of 128 Hz and its first four harmonics (256 Hz, 384 Hz, 512 Hz and 640 Hz) are aligned across the HCQT channels.
This allows a series of local convolutions to capture relationships across frequencies that would require much larger receptive fields when using, for example, a single CQT.

\subsection{Output Representation}
We train our models to produce an ideal pitch salience representation: a time-frequency representation where all bins corresponding to a $f_0$ have a value of 1, and all other bins have a value of 0.
In this work, we use a grid with a time resolution of $11.6$ ms per time frame and a frequency resolution of 5 bins per semitone. 

In practice, $f_0$ annotations are not always accurate up to one fifth of a semitone, and we do not want to penalize our models heavily for estimates that are within a quarter-tone of the annotated label.
We achieve this by adding Gaussian blur in frequency to the binary salience matrix, assigning decaying positive values to frequencies within 2 bins of the annotated frequency.

\subsection{Architectures}
\begin{figure*}
\begin{center}
\includegraphics[width=2\columnwidth]{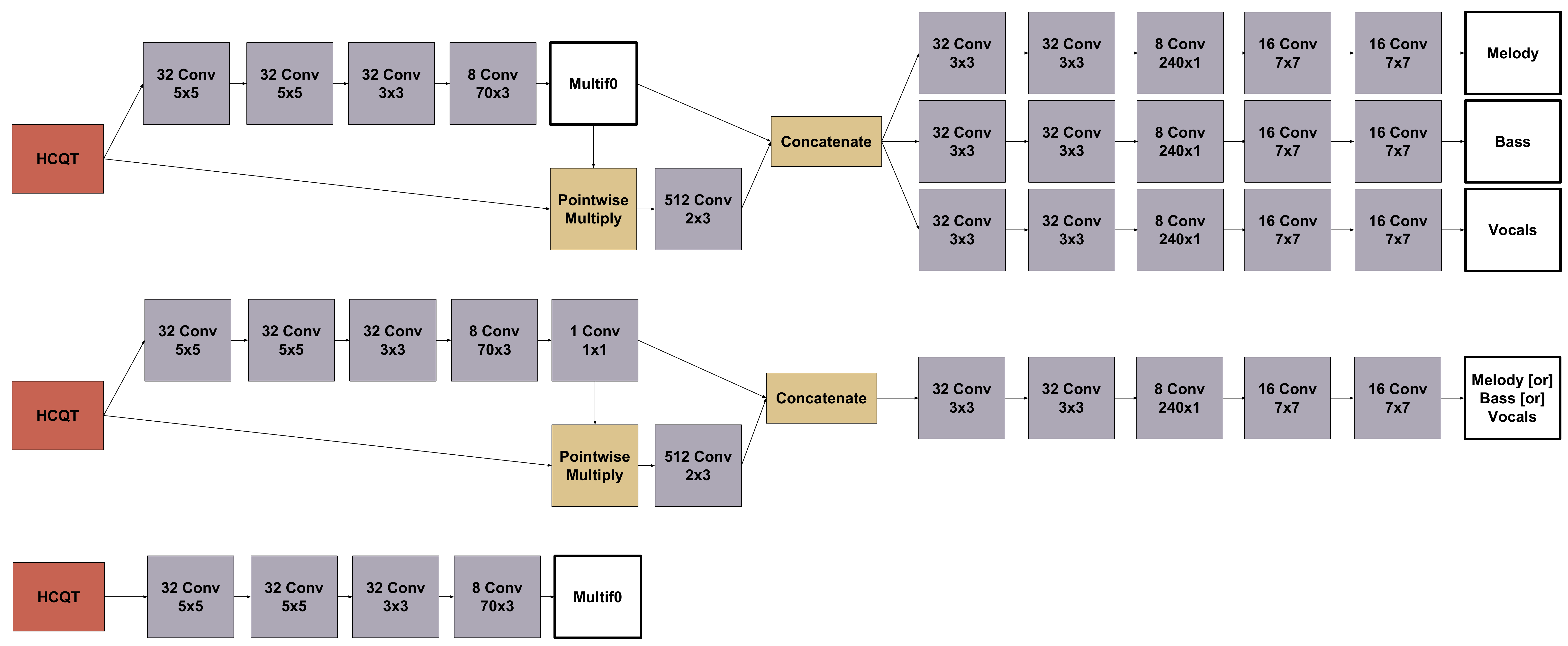}
\caption{(Top) Multitask architecture producing multiple-$f_0$, melody, bass and vocal salience outputs. (Middle) Single-task architecture producing Melody (or Bass or Vocal) salience output. (Bottom) Single-task architecture producing multiple-$f_0$ output. Red boxes represent HCQT input features, grey boxes denote convolution operations, yellow boxes denote other mathematical operations, and white boxes are used for output layers from which loss is measured.}\label{fig:mtask_arch}
\end{center}
\end{figure*}

We design a multitask architecture (Figure~\ref{fig:mtask_arch} (top)) that uses the multiple-$f_0$ estimation output to improve the estimations for the auxiliary tasks, and at the same time allows the estimates from the auxiliary tasks to reinforce the multiple-$f_0$ estimation because of the additional training data.

The input HCQT is passed through four convolutional layers and produces a multiple-$f_0$ salience estimate $M[t, f]$ as the first output.
To enforce the multiple-$f_0$ superset relationship, we use the multiple-$f_0$ salience estimate to mask each channel of the HCQT: 
\begin{equation}
\mathcal{H}_{\text{masked}}[h, t, f] = \mathcal{H}[h, t, f] \times M[t, f]
\end{equation}
where $M$ is the matrix of predicted multiple-$f_0$ values, and $M[t, f]$ is the value at time $t$ and frequency $f$ with a value ranging from 0 to 1.
This in essence only allows $f_0$ values detected in the multiple-$f_0$ salience estimate to be active in the other output layers.
The masked HCQT features are passed through a single convolutional layer designed to capture local timbre information (weighted combinations of harmonic information over a short time period), and these feature maps are concatenated with the multiple-$f_0$ salience estimates.
Including timbre information is necessary to distinguish $f_0$ values corresponding to specific sources, particularly vocals~\cite{bosch2017phd,dzhambazov2016singing}.

The concatenated features are passed through three identical sub-networks, each with 5 convolutions. 
They output melody, bass, and vocal salience estimates respectively.
One of the convolutional layers in these sub-networks is ``tall'' to capture tasks-specific frequency relationships, e.g. distinguishing low from high frequency content in bass detection.
Rectified linear activations followed by batch normalization are applied after each convolution layer except the last which has sigmoid activations.

Figure~\ref{fig:mtask_arch} (Middle) shows a single-task architecture we use for comparison in our subsequent experiments.
Note that the ``Multiple-$f_0$'' output layer is replaced by a single $1 \times 1$ convolution, to make the architectures equivalent.
All auxiliary single-task architectures are identical, and the data used for training determines what is produced by the output layer.
The multiple-$f_0$ single task architecture in Figure~\ref{fig:mtask_arch} (Bottom) is identical to the multitask architecture, but without the sub-networks for the other tasks.

\subsection{Training}

All models in the subsequent experiments are trained using the data described in Section~\ref{sec:training-data}.
In this work we use a cross-entropy loss function:
\begin{equation}\label{eq:loss}
L(y, \hat{y}) = -y\log(\hat{y}) - (1 - y)\log(1-\hat{y})
\end{equation}
which is computed point-wise over the estimated and target salience matrices $\hat{Y}$ and $Y$ respectively.
We will use $L(Y, \hat{Y})$ to denote the average loss over all point wise pairs in $Y$ and $\hat{Y}$:
\begin{equation}
L(Y, \hat{Y}) = \frac{1}{TF}\sum_{t=0}^{T-1} \sum_{f=0}^{F-1} (L(Y[t, f], \hat{Y}[t, f]))
\end{equation}
To train a multitask model, we generate feature/label tuples $\left(X, Y_1, \dots, Y_n \right)$ where $X$ is the input feature matrix and each $Y_i$ corresponds to the salience matrix for a particular task (e.g. melody).
We train the model to minimize the sum of the loss functions for each of the $n$ tasks:
\begin{equation}\label{eq:loss-sum-weight}
L_{\text{overall}} = \sum_{i=1}^{n} w_i \cdot L(Y_i, \hat{Y}_i)
\end{equation}
For a given $X$ we may not have labeled data for all tasks, i.e. some subset of $(Y_1, \dots, Y_n)$ is not available.
To compensate for this, we apply weights $w_i$ to individual losses, setting $w_i=1$ by default, and setting $w_i = 0$ whenever $Y_i$ is unavailable.

Training is performed in batches, and each batch contains at least one example from each task.
Training examples are sampled randomly from the dataset, and the sample's time window (50 frames $\approx 0.58$ s) from within the audio track is chosen randomly as well.
We train with a batch size of 4, and compute the validation error over 50 batches once every epoch of 50 training batches.
We perform early stopping if the validation error does not improve after 25 epochs.

The model is implemented using the Keras~\cite{chollet2015keras} library with TensorFlow~\cite{abadi2016tensorflow} and optimized using the Adam~\cite{kingma2014adam} optimizer.
Implementations of the models described in this work are made available online.\footnote{\url{https://github.com/rabitt/multitask-f0}}

\section{Training Data}\label{sec:training-data}

Datasets containing musical audio and corresponding $f_0$ labels are quite scarce, primarily because creating human-annotated $f_0$ labels is labor intensive and difficult~\cite{bittner2014medleydb}.
For all tasks except melody estimation, the largest datasets contain only a single type of music such as piano solos (multiple-$f_0$), pop karaoke (vocal $f_0$) or jazz ensembles (bass $f_0$).
In order for our model to generalize to a non-specific musical context, we would like each task to be trained on a variety of musical genres.
Furthermore, there is no single dataset containing annotations for all four of the tasks we consider.
For these reasons, we rely on semi-automatically labeled $f_0$ data, created by remixing a diverse set of \emph{multitrack} audio data.

In this work, we use multitracks from the MedleyDB~\cite{bittner2014medleydb} dataset, as well as a collection of commercial multitracks, resulting in a dataset of 320 multitracks.
For the purposes of our experiments we split these 320 tracks into artist-conditional training, validation, and test groups -- i.e. tracks by the same artist may only appear in one group.
We constrain the test group to only contain tracks from the publicly available MedleyDB dataset.
The resulting training, validation, and test groups contain 264, 28, and 28 original tracks respectively.

The unique advantage of multitrack data is that it allows for countless types of remixes and recombinations of the stems, giving us control over the properties of the resulting mixtures.
For our application, we focus on augmentations that allow us to increase the amount of audio-annotation pairs for different $f_0$ estimation tasks.

In particular, our goal is to create remixes of a song where the $f_0$ content of every stem is known.
To do this we break instruments into 3 categories: \emph{monophonic} (instruments that typically play a single pitched line), \emph{polyphonic} (instruments that play multiple pitches at once such as piano or guitar), and \emph{unpitched} (instruments that do not play pitched content, such as percussion).

For each multitrack we use 4 different types of mixtures:
\begin{itemize}
\item \texttt{ORIG}: Original mix (320 tracks)
\item \texttt{RMX-PG}: Remix containing unpitched, monophonic, synthesized keyboard and synthesized guitar stems (227 tracks)
\item \texttt{RMX-P}: Remix containing unpitched, monophonic and synthesized keyboard stems (225 tracks)
\item \texttt{RMX}: Remix containing unpitched and monophonic stems (222 tracks)
\end{itemize}

\begin{figure}[ht]
\begin{center}
\includegraphics[width=\columnwidth]{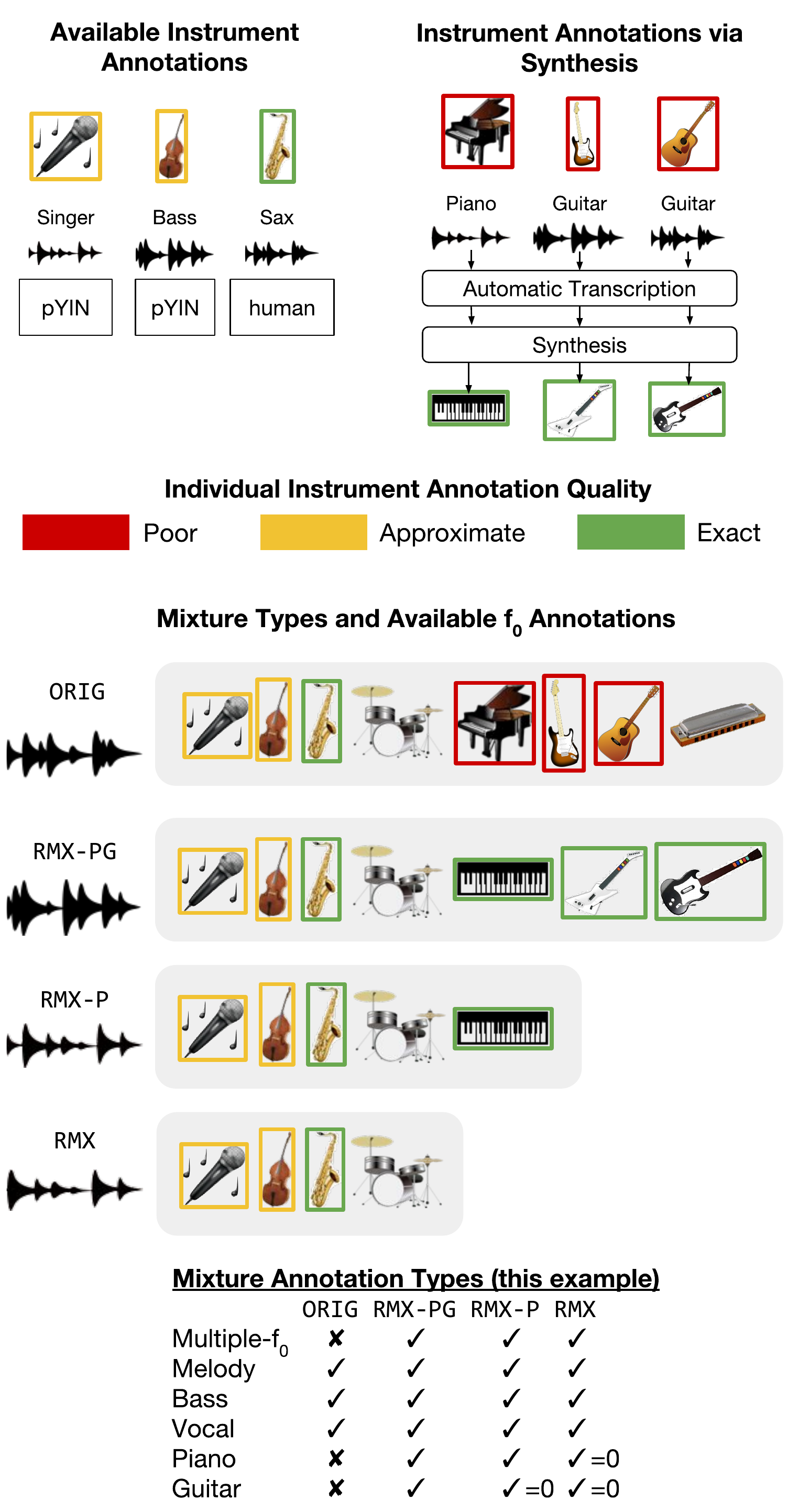}
\caption{Diagram illustrating each type of mixture and corresponding annotations for an example multitrack. (Top left) Available $f_0$ annotations that are of sufficient quality to train on. In this example, the saxophone plays the melody. (Top right) Synthesis process for piano and guitar stems --- stems are transcribed and the transcription is synthesized to create a new stem. (Middle) Instruments present in each type of mixture. Each instrument is highlighted based on the quality of the available annotations. (Bottom) Table of available mixture-level annotations available for training in this example. Guitar annotations are available e.g. for \texttt{RMG-P} because we know there is no guitar present, thus the annotation is empty.}\label{fig:data}
\end{center}
\end{figure}

Figure~\ref{fig:data} illustrates these four types of mixes and their available types of annotations for a particular example.
Our goal is to use or create realistic musical mixtures with as many annotation types as possible.
In this example, the original mix \texttt{ORIG} has 3 monophonic instruments (singer, bass, and saxophone), 1 unpitched instrument (drums), and 4 polyphonic instruments (piano, electric guitar, acoustic guitar, and harmonica).
For each instrument, we have different qualities of $f_0$ annotations available, which we combine to create mixture annotations.
The only available human-labeled $f_0$ annotation annotation is for the saxophone (the melody in this example).
We generate approximate $f_0$ annotations for the other monophonic instruments by running a monophonic pitch tracker on the instrument stems, described in detail in Section~\ref{sec:synth-f0}.
For the piano and guitar stems, the automatic transcriptions produce poor annotations, and are not accurate enough to use in training.
The only mixture-level annotations we can create in this example for \texttt{ORIG} are melody, bass and vocals; we cannot create a multiple-$f_0$ annotation for \texttt{ORIG} because we do not accurately know the $f_0$ information played by any of the polyphonic instruments.
However, we can use the transcriptions of the polyphonic stems to generate synthesized stems for which we do know the exact $f_0$ information, which we later use in remixes (see Sections~\ref{sec:piano-rep} and~\ref{sec:guitar-rep}).
When we remove or replace the polyphonic stems as in \texttt{RMX-PG}, \texttt{RMX-P} and \texttt{RMX}, we create mixtures for which we know all of the $f_0$ information for all pitched instruments (see Section~\ref{sec:artifical-mix}).
Note that in this work we only create synthesized stems for piano and guitar, and remove all other polyphonic instruments (harmonica in this example).

Table~\ref{tab:task-totals} gives the number of tracks available with annotations for each task and mixture type.
While there are 320 multitracks available total, only 227 of those have ``clean'' stems (without bleed), so we only create remixes for these 227.
There are two tracks with only guitar and percussion, and three with only keyboard instruments and percussion, which is why the numbers differ slightly across remix types.
It is important to note that some of the numbers in Table~\ref{tab:task-totals} are deceptively high --- when a task is known to have no $f_0$ content (e.g. a track with no bass), the annotations exist, but are empty, resulting in a salience representation that is only zeros.
We refer to these annotations as ``null'' annotations.
Table~\ref{tab:annotation-nums} shows the number of null annotations per task and mixture type.

\begin{table}
\begin{center}
\begin{tabular}{ c c c c c c }
\hline 
\textbf{Task} & \texttt{ORIG} & \texttt{RMX-PG} & \texttt{RMX-P} & \texttt{RMX} & \textbf{Total} \\ 
\hline
Multiple-$f_0$ & 0 & 227 & 225 & 222 & \textbf{674} \\
Melody & 153 & $\mid$ & $\mid$ & $\mid$ & \textbf{827} \\
Bass & 244 & $\mid$ & $\mid$ & $\mid$ & \textbf{918} \\
Vocals & 183 & $\mid$ &$\mid$ & $\mid$ & \textbf{857} \\
Piano & 229 & $\mid$ & $\mid$ & $\mid$ & \textbf{903} \\
Guitar & 122 & $\mid$ & $\mid$ & $\mid$ & \textbf{796} \\
\hline
\textit{Total Available} & \textit{320} & \textit{227} & \textit{225} & \textit{222} & \textit{994}\\
\hline
\end{tabular}
\end{center}
\caption{Number of annotated tracks per task and mixture types. Vertical bars indicate that the value is the same for the entire column. Note that for many of these annotations, the ground truth information is that there is no $f_0$ content for a task (i.e. a ``null'' annotation). See Table~\ref{tab:annotation-nums} for the number of null annotations for each task and mixture type.}
\label{tab:task-totals}
\end{table}

\begin{table}
\begin{center}
\begin{tabular}{ l  llll  l }
\hline 
\textbf{Task} & \texttt{ORIG} & \texttt{RMX-PG} & \texttt{RMX-P} & \texttt{RMX} & \textbf{Total}\\ 
\hline
\textbf{Multiple-}$f_0$ & 0 & 3 & 4 & 5 & \textbf{12} \\
\textbf{Melody} & 15 & 48 & 46 & 43 & \textbf{152} \\
\textbf{Bass} & 77 & 58 & 56 & 53 & \textbf{244} \\
\textbf{Vocals} & 110 & 70 & 68 & 65 & \textbf{313} \\
\textbf{Piano} & 229 & 182 & 180 & 222 & \textbf{813} \\
\textbf{Guitar} & 122 & 79 & 225 & 222 & \textbf{648} \\
\hline 
\end{tabular}
\end{center}
\caption{Number of null annotations per task and mixture type. A null annotation refers to one where we know that the task is not present -- e.g. a track with no vocals will have a null vocal annotation.
A multiple-$f_0$ annotation is null if a mixture contains only unpitched instruments.}
\label{tab:annotation-nums}
\end{table}

In the following sections, we discuss the details of how these mixes are created, and how $f_0$ estimations are created for each type of instrument we consider.

\subsection{Generated Mixes}\label{sec:artifical-mix}
To create carefully controlled remixes, we both remove and replace stems from multitracks.
Given a new set of stems, we want to re-create mixtures that are realistic, and to do that we make the mixtures as close as we can to the original mixture.
Even when using the original stems as source material, a simple linear sum of the stems will not necessarily be a good approximation of the original mix: the stems may not be the same volume as they occur in the mix, and the overall mix could have mastering effects such as compression or equalization applied.
In this work, we ignore mastering effects, and model the mix $y[n]$ (containing N samples) as a linear combination of the stems $x_1, x_2, \ldots, x_M$:
\begin{equation}
y[n] \approx \sum\limits_{i=1}^M a_i x_i[n]
\end{equation}
The weights $a_i$ are approximated by minimizing a non-negative least squares objective $||X\mathbf{a} - Y||_2$ over $\mathbf{a}$ for $a_i > 0$, where $X$ is the $N \times M$ matrix of the absolute values of the stem audio signals $|x_i[n]|$, $\mathbf{a}$ is the $M \times 1$ vector of mixing weights $a_i$, and $Y$  the $N \times 1$ is the absolute value of the mixture audio signal $|y[n]|$.

Our training mixes optionally remove and/or replace a subset of the original stems.
Let $x_1, x_2, \ldots, x_M$ be the original stems, $\tilde{x}_1,\ldots,\tilde{x}_k,$ be replacement stems, and $x_{k+1}, \ldots x_{k+K}$ ($K \le M$) be removed.
The resulting mix $\tilde{y}$ is created by simply substituting $x_1,\ldots,x_k$ with $\tilde{x}_1,\ldots,\tilde{x}_k$, scaled by the approximated mixing weights $a_i$, and excluding the removed stems:
\begin{equation}
\tilde{y}[n] = \sum\limits_{i=1}^k a_i\tilde{x}_i[n] + \sum\limits_{i=k+K+1}^{M} a_i x_i[n]
\end{equation}
Note that this formulation assumes that replaced stems have the same amplitude envelope as the original stems so that the mixing weight corresponding to the original stem is also meaningful for the replaced stem.


\subsection{Single-$f_0$ Annotations}\label{sec:synth-f0}
We use the pYIN pitch tracking algorithm~\cite{Mauch:2014gm} to estimate the ground truth pitch annotations for solo, monophonic instrument recordings.
Monophonic instruments commonly used in this way include male and female vocals, violin, saxophone, and flute.
Instruments that often play multiple notes simultaneously such as guitars and pianos are not considered here.
While we found pYIN to be insufficient for some instrumental sources such as electric guitar, it is quite accurate for the majority of monophonic sources.

pYIN is run with hop size 256 and window size 8192 at a sample rate of 44100.
We post-process the output of pYIN by removing any estimated pitches that occur when the source is not estimated as active~\cite{bittner2014medleydb}.
We measure the accuracy of these estimated $f_0$ outputs by comparing them against the 96 human-labeled annotations in MedleyDB (original release), and found they had an \texttt{RPA} (Eq.~\ref{eq:RPA}) of 80\% with an \texttt{OA} (Eq.~\ref{eq:OA}) of 79\%.

\subsection{Piano Replacement}\label{sec:piano-rep}
Piano transcription estimates are generally far less reliable than pitch tracker estimates for monophonic instruments.
Instead of simply using the output of a piano transcription as a rough proxy for the ground truth $f_0$ as we did in Section~\ref{sec:synth-f0}, we extend the idea proposed in~\cite{salamon2017automatic}.
Specifically, we use the output of a piano transcription algorithm to drive a simple piano synthesizer to create a version of the original piano stem that is similar to the original, and for which we have a perfect $f_0$ transcription.

We use the transcription algorithm  proposed in~\cite{benetos2015efficient}, which estimates a sequence of note events of the form (start time, end time, MIDI note), which we then convert to MIDI-format.
Each MIDI-note velocity is estimated as the average log-normalized energy of a semitone-resolution CQT within the time-frequency region of the note.
The resulting MIDI file is rendered to audio using Fluidsynth~\footnote{\url{http://www.fluidsynth.org/}} using a selected sound font.
The audio file is used as a replacement stem, and the MIDI file is used to generate an $f_0$ transcription.

\subsection{Sound Font Selection}\label{sec:sound-font}
Our goal is to find a sound font that matches the timbre of the original audio file as closely as possible.
For each of the considered instruments -- piano, electric piano, and synthesizer -- we manually curate a bank of sound fonts corresponding to the instrument timbre.
We select sound fonts such that there is an exact correspondence between the MIDI notes given and the sounding pitches in the audio file; we found that many sound fonts in the wild surprisingly do not fit this property.

Given a sound font bank and an original piano / electric piano / synthesizer stem, we select the best available font using the following procedure.
We render a simple MIDI file (containing scales and chords) to audio using each sound font in the bank.
For each of these audio files we compute 40-dimensional MFCCs for the full audio file and then compute the average.
The resulting set of vectors is standardized across all fonts so that each dimension is 0-mean unit variance, resulting in a set of 40-dimensional feature vectors $\{\mathbf{s}_1, \mathbf{s}_2, ... \mathbf{s}_n\}$ where $n$ is the number of sound fonts.
In the same fashion, we compute the normalized, averaged 40-dimensional MFCC query vector $\mathbf{q}$ from the original audio.
Finally we choose the sound font with the smallest Euclidean distance between the query vector and the sound font feature vector:
\begin{equation}
\argmin_i || \mathbf{q} - \mathbf{s}_i ||_2
\end{equation}

\subsection{Guitar Replacement}\label{sec:guitar-rep}
We follow a similar process for guitar as for piano -- given a MIDI file corresponding roughly to the $f_0$ content in a solo guitar stem, we render the MIDI file to audio using Fluidsynth with the sound font matching procedure described in Section~\ref{sec:sound-font}.
However, the available transcription methods do not work as well for guitar as they do for piano, and MIDI renderings of their output are far from the sound of the original audio file.
In an effort to create a more ``realistic'' synthesis, we employ a different transcription approach for guitar, aimed at recreating \emph{strummed chords}.

We first run the chord recognition algorithm described in~\cite{mcfee2017_structured} over the solo guitar stem, resulting in sequences of \texttt{(start time, end time, chord label)}.
Next we compute onsets and offsets from the solo guitar stem using a standard onset detection algorithm~\cite{bock2013maximum}.
We consider each onset the start of a guitar \emph{strum}, and the following offset the release time.
For our purposes, we define a strum to be a sequence of notes where the onset times of each note occurs slightly after the previous note.

To simulate a strum, for a \texttt{(onset time $t_{\text{onset}}$, offset time $t_{\text{offset}}$, chord label)} group we first select a chord voicing (set of notes in the chord) from a dictionary\footnote{The chord voicing dictionary was obtained from a corpus of online guitar tabs~\cite{humphrey2014music}.} of plausible chord labels -- guitar voicing pairs, which results in an ordered sequence of MIDI note numbers $(m_1, \cdots, m_I)$ where $I$ is the number of notes in the voiced chord.
When more than one voicing is available for a given chord label, we choose the voicing that has the smallest MIDI note number edit-distance from the previous chord voicing.
Strums alternate between ``down'' (i.e. in voicing order), and ``up'' (in reverse voicing order).
For $m_1$ the start time $t_{s_1} = t_{\text{onset}} - 0.01$.
For subsequent $m_i$, the start time $t_{s_{i-1}} = t_{s_i} + \delta$ where $\delta$ is random and chosen uniformly between 0.01 s and 0.05 s.
For all $m_i$ the end time is $t_{\text{offset}}$.

The rendered audio file is used as a replacement stem for the original guitar stem, and the MIDI note events are used to generate an exact $f_0$ transcription of the replacement guitar.

\subsection{Limitations}\label{sec:limitations}
While our approach to generating $f_0$ annotations allows us to vastly increase the size of our training set, the data is limited in how well it represents real music.
The estimated single-$f_0$ annotations often contain mistakes, causing a mismatch between the stem and its annotation.
The mistakes can be particularly severe for bass lines.

For piano and especially guitar, we rely on transcriptions to create replacement stems, but in both cases the transcriptions are often far from the originals.
This is especially true for guitar stems that do not follow a strumming pattern but instead play a melodic line or a finger picking pattern -- cases where our method of transcribing strums breaks down.
While there is a perfect correspondence between the guitar/piano transcriptions and their replacement stems, the mistakes can cause the resulting mixes to be musically incoherent.

The sound fonts used for piano and guitar are limited, and in particular guitar sound fonts can be unrealistic and fail to match the original guitar sound.
This potentially risks over-fitting to specific sound fonts and not generalizing well to natural guitar/piano.

Finally, our dataset is skewed towards pop and rock.
In general, there is very little multitrack data available for genres such as classical music, and as a result our dataset is unrepresentative of those styles.


\section{Experimental Setup}\label{sec:experiments-setup}

\subsection{Evaluation Data}
We evaluate our models on the most comprehensive public datasets available for each $f_0$ estimation task.
We outline and give a brief description of each of these datasets below.

\noindent
\textbf{Multiple-$f_0$}
\begin{itemize}
\item \textbf{Bach10}~\cite{duan2010multiple}: ten 30-second recordings of a quartet performing Bach chorales.
\item \textbf{Su}~\cite{su2015escaping}: 10 excerpts of real-world classical recordings, including examples of piano solos, piano quintets, and violin sonatas.
\item \textbf{MAPS}~\cite{emiya2008transcription}: 238 fully annotated solo piano recordings of classical pieces in different recording conditions.
\item \textbf{MDB$_{mf0}$} The 28 tracks (\texttt{RMX}) from the MedleyDB test group.
\end{itemize}

\noindent
\textbf{Melody}
\begin{itemize}
\item \textbf{Orchset}~\cite{bosch2016evaluation}: 64 full orchestral excerpts with melody annotations.
\item \textbf{WJ}$_{mel}$ The Weimar Jazz database - Melody~\cite{pfleiderer2010jazzomat}: 500 jazz recordings with main melody annotations.
\item \textbf{MDB$_{mel}$} The 28 tracks (\texttt{ORIG}) from the MedleyDB test group.
\end{itemize}

\noindent
\textbf{Bass}
\begin{itemize}
\item \textbf{WJ}$_{bass}$ The Weimar Jazz database - Bass~\cite{pfleiderer2010jazzomat}: 40 jazz excerpts with bass line annotations.
\end{itemize}

\noindent
\textbf{Vocal-$f_0$}
\begin{itemize}
\item \textbf{Ikala}~\cite{chan2015vocal}: 252 excerpts of Chinese karaoke
\end{itemize}

\subsection{Metrics}
We evaluate the accuracy of our systems using the standard melody extraction (for melody, bass and vocal-$f_0$) and multiple-$f_0$ evaluation metrics.

\subsubsection{Single-$f_0$}
The five standard evaluation metrics for melody evaluation are presented in detail in~\cite{Salamon:2014fh} and outlined here.
Let $\mathcal{T}(a)$ be a frequency comparison function, such that it returns 1 if the frequencies are within a quartertone and 0 if they are not:
\begin{equation}
\mathcal{T}(a) = \mathds{1}_{|a| < 0.5}
\end{equation}
Let $f[n]$ and $v[n]$ be the true values of the frequency and voicing vectors, and $\hat{f}[n]$ and $\hat{v}[n]$ be the estimates.
$v[n]$ and $\hat{v}[n]$ are binary vectors, and $f[n]$ and $\hat{f}[n]$ are vectors of frequencies in units of Hz.
Let $f_{\mathcal{M}}[n]$ and $\hat{f}_{\mathcal{M}}[n]$ give the values of the frequency along a log frequency scale where $f_{\mathcal{M}}[n]$ is related to $f[n]$ by:
\begin{equation}
f_{\mathcal{M}}[n] = 12\log_2\left(\frac{f}{f_{\text{ref}}} \right)
\end{equation}
Raw pitch accuracy (\texttt{RPA}) measures the percentage of time frames where a pitch is present in the reference where the estimate is correct within a threshold (ignoring voicing mistakes). 
\begin{equation}\label{eq:RPA}
\texttt{RPA} = \dfrac{\sum\limits_{n=0}^{N-1} v[n] \mathcal{T}\left(\hat{f}_{\mathcal{M}}[n] - f_{\mathcal{M}}[n] \right)}{\sum\limits_{n=0}^{N-1} v[n]}
\end{equation}
Raw chroma accuracy (\texttt{RCA}) is the same as raw pitch accuracy, except that octave errors are forgiven by mapping all frequencies onto a single octave.
\begin{equation}\label{eq:RCA}
\texttt{RCA} = \dfrac{\sum\limits_{n=0}^{N-1} v[n] \mathcal{T}\left(\langle \hat{f}_{\mathcal{M}}[n] - f_{\mathcal{M}}[n] \rangle_{12} \right)}{\sum\limits_{n=0}^{N-1} v[n]}
\end{equation}
where
\begin{equation}
\langle a \rangle_{12} = a - 12 \left\lfloor \frac{a}{12} + 0.5 \right\rfloor .
\end{equation}
Raw chroma accuracy is at least equal to raw pitch accuracy.
Voicing recall (\texttt{VR}) measures the number of correctly predicted voiced frames over the total number of voiced frames, and the voicing false alarm rate (\texttt{VFA}) measures the percentage of frames that are estimated as voiced but are actually unvoiced.
\begin{equation}\label{eq:VR}
\texttt{VR} = \dfrac{\sum\limits_{n=0}^{N-1} \hat{v}[n]v[n]}{\sum\limits_{n=0}^{N-1} v[n]}
\end{equation}
\begin{equation}\label{eq:VFA}
\texttt{VFA} = \dfrac{\sum\limits_{n=0}^{N-1} \hat{v}[n](1 - v[n])}{\sum\limits_{n=0}^{N-1} (1 - v[n])}
\end{equation}
Overall accuracy (\texttt{OA}) combines the voicing accuracy and the raw pitch accuracy giving the percentage of correctly predicted frames.
{\small
\begin{equation}\label{eq:OA}
\texttt{OA} = \frac{1}{N}\sum\limits_{n=0}^{N-1} v[n] \mathcal{T}\left(\hat{f}_{\mathcal{M}}[n] - f_{\mathcal{M}}[n]  \right) + (1-v[n])(1-\hat{v}[n])
\end{equation}
}

\subsubsection{Multiple-$f_0$}
Multiple-$f_0$ systems have historically been evaluated using a set of metrics defined in~\cite{poliner2007discriminative}.
The ground truth value $f[n]$ and estimate $\hat{f}[n]$ at frame $n$ can each have multiple pitch values, denoting the pitches of all active sources in that frame.
The number of values per frame varies with time, and we denote $c[n]$ as the number of pitch values in $f[n]$, and $\hat{c}[n]$ as the number of pitch values in $\hat{f}[n]$.
For a given frame $n$, let $\texttt{TP}[n]$ be the number of correctly transcribed pitches, $\texttt{FP}[n]$ be the number of pitches present in $\hat{f}[n]$ that are not present in $f[n]$, and let $\texttt{FN}[n]$ be the number of pitches present in $f[n]$ that are not present in $\hat{f}[n]$.

The accuracy (\texttt{Acc}) metric is defined as the sum over frame level accuracies:
\begin{equation}
	\texttt{Acc} = \dfrac{\sum\limits_{n=0}^{N-1} \texttt{TP}[n]}{\sum_{n=0}^{N-1} \texttt{TP}[n] + \texttt{FP}[n] + \texttt{FN}[n]}
\end{equation}
which gives a value between 0 and 1, where 1 corresponds to a perfect transcription.

\section{Experiments}\label{sec:experiments}

In this section we present experiments to address the following questions:
\begin{itemize}
    \item Does a multitask architecture outperform equivalent single-task architectures? (Section~\ref{sec:exp-mtaskvstask})
    \item Ablation studies: (Section~\ref{sec:ablation})
    \begin{enumerate}
        \item Does performance improve with each additional task? (Section~\ref{sec:ntasks})
        \item Does the addition of synthesized instruments in our training data help or hurt performance? (Section~\ref{sec:synth-data})
        \item Does the HQCT provide an improvement over a standard CQT? (Section~\ref{sec:exp-hcqt})
    \end{enumerate}
    \item How well does the multitask model compare with the current state of the art approaches? (Section~\ref{sec:exp-baselines})
\end{itemize}

In each of the following experiments, we compare our multitask model (Figure~\ref{fig:mtask_arch}, Top) consisting of four tasks against other approaches and variants of our model.
In every plot, the results for the multitask model are colored in dark blue.
Each plot in this section gives results as boxplots - metrics are computed per track, and the boxplots indicate the mean, median, quartiles, and extrema of the per-track metrics.

\subsection{Multitask versus Single-task Architectures}\label{sec:exp-mtaskvstask}
In these experiments we examine the effectiveness of the multitask architecture compared with equivalent single-task architectures.
First we examine the four-task multitask architecture (multiple-$f_0$, melody, bass, and vocal-$f_0$ as in Figure~\ref{fig:mtask_arch}, top) against four single-task architectures, sown in Figure~\ref{fig:mtask_arch}, middle and bottom.
Unless otherwise noted, each of the models are trained on the same set of audio.

\begin{figure}
\begin{center}
\includegraphics[width=\columnwidth]{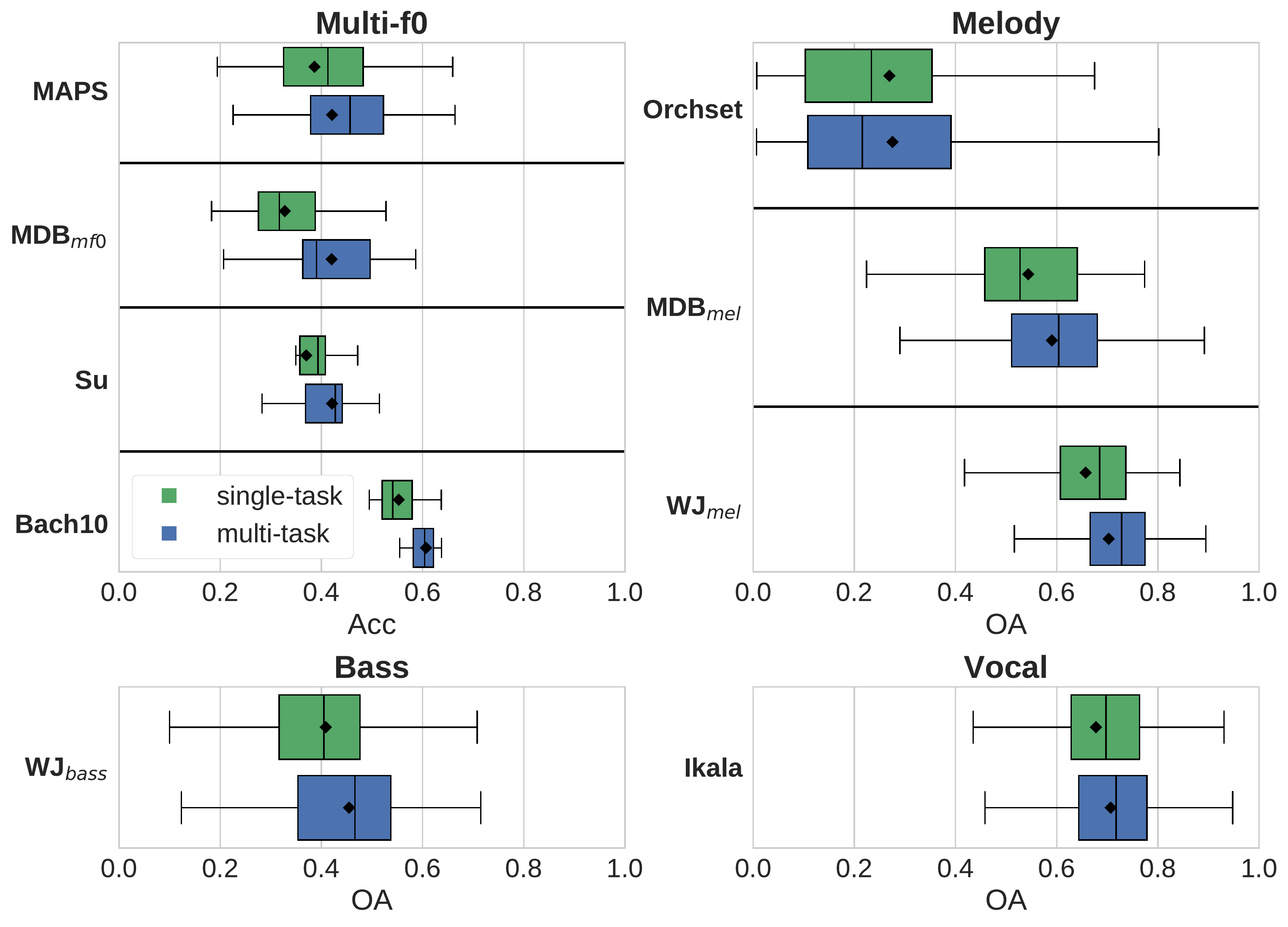}
\end{center}
\caption{Multitask vs single-task results. The multitask results (blue) are produced by one multitask model that produces four outputs. The single task results (green) are produced by four different single-task models, each trained to optimize a different task.}\label{fig:mtask-v_stask}
\end{figure}

Figure~\ref{fig:mtask-v_stask} shows the overall scores for the multitask versus single-task architectures on each dataset.
We see that, in accordance with results from other domains, the multitask architecture outperforms its single-task equivalents for each task on each dataset.
The performance difference is greater for the multiple-$f_0$ outputs than for the subtasks, likely because the multiple-$f_0$ output is reinforced by all other tasks, while melody, bass and vocal-$f_0$ are only directly reinforced by the multiple-$f_0$ output.

\begin{figure*}
\begin{center}
\includegraphics[width=2\columnwidth]{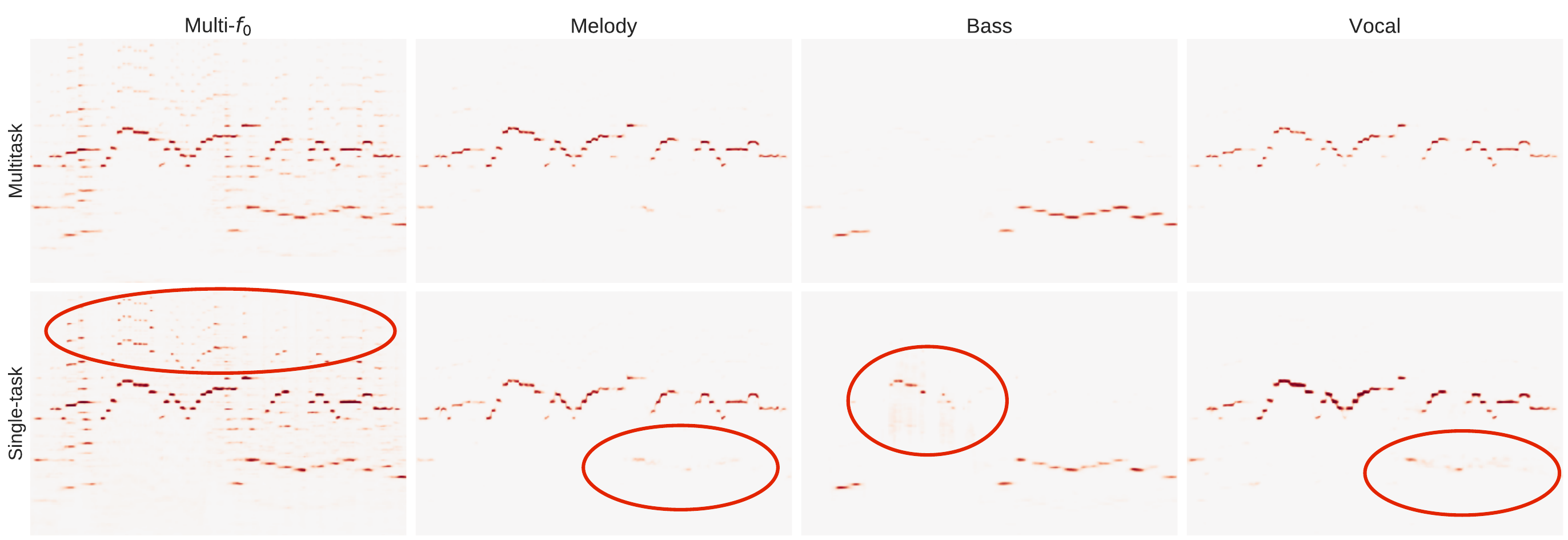}
\end{center}
\caption{Example of multitask versus single-task outputs on a Jazz excerpt from Clifford Brown's ``Sandu'' (from \textbf{WJ}$_{bass}$). The excerpt is taken from the first 10 seconds of the trumpet solo. In each plot the frequency axis (y-axis) is shown on a log-scale, and ranges from 32.7 Hz (C1) to 2093 Hz (C6). The color axis shows the predicted salience values, with white representing a value of 0 and darker shades of red representing values closer to 1. In this example, the multitask model has an \texttt{OA} of 0.75 for melody and 0.57 for bass. The single-task melody model has an \texttt{OA} of 0.69, and the single-task bass model has an \texttt{OA} of 0.48. Note that we do not have ground truth annotations for multiple-$f_0$ on this example. A selection of false alarm errors are circled in red.}\label{fig:mtask-v_stask-qual}
\end{figure*}

Figure~\ref{fig:mtask-v_stask-qual} shows an output produced for the same example using the multitask and single-task architectures.
In the multiple-$f_0$ predictions, the biggest difference between the multitask and the single-task outputs is that the multitask output is better at de-emphasizing the harmonics produced by the melodic instrument (trumpet).
Both the multitask and single-task models are equally good at isolating the melodic line, but since the multiple-$f_0$ layer shares weights with the melodic layer in the multitask layer, the multiple-$f_0$ layer is better able to distinguish harmonics from the fundamental.
For the bass output, in the single-task model, part of the melody is mistaken as the bass line, and overall the correctly identified bass notes are less salient than in the multitask model.
The excerpt shown in Figure~\ref{fig:mtask-v_stask-qual} does not contain vocals, yet both models predict a vocal line that is more or less equivalent to the melodic line, though the multitask model predicts a less salient vocal line than the single-task model.
This suggests that neither model is actually discriminating information based on timbre, or that there is not enough timbre information available to the model (e.g. not enough harmonics).
Instead, the model appears to be identifying vocals by frequency position and by rapidly changing $f_0$ trajectories (similar to the findings of~\cite{schluter2016learning}).

\subsection{Ablation Studies}\label{sec:ablation}

In the following experiments, we explore the effect of specific choices made in the design of our model.

\subsubsection{Performance vs. Number of Tasks}\label{sec:ntasks}

\begin{figure}
\begin{center}
\includegraphics[width=\columnwidth]{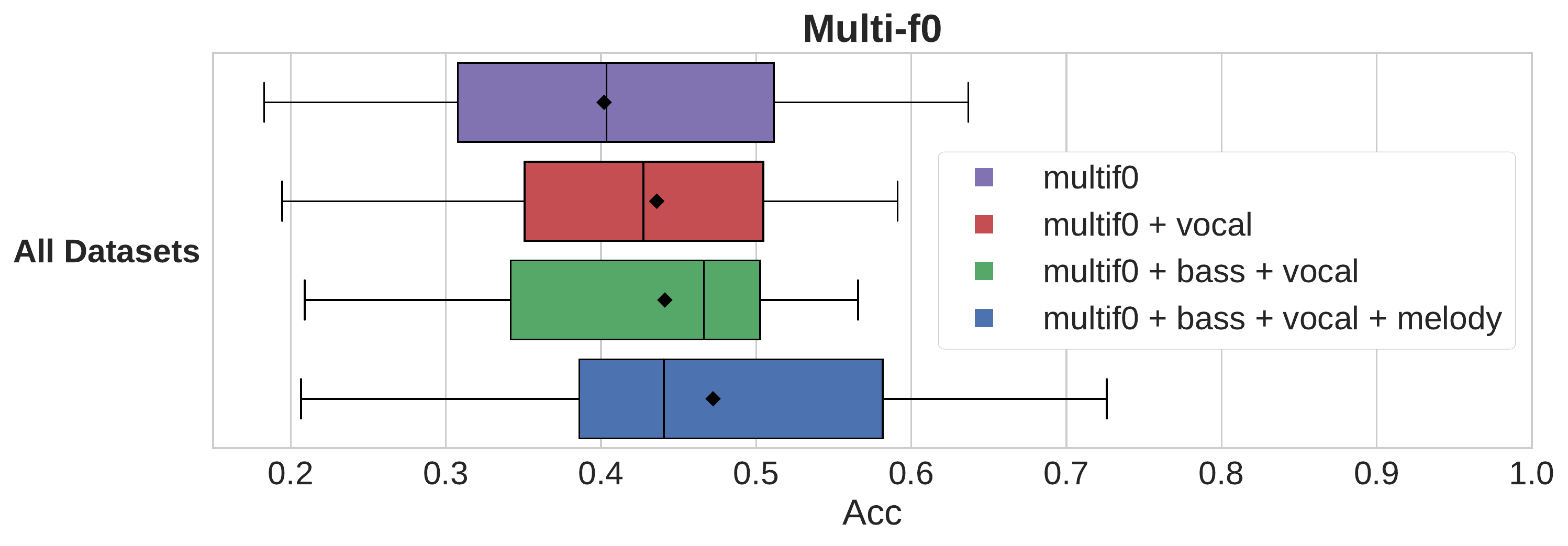}
\end{center}
\caption{The overall influence on multiple-$f_0$ performance from adding more tasks to the multitask model. Each model is trained on the same amount of audio data - the only difference is the number of outputs which receive a loss signal.}\label{fig:mtask-combos}
\end{figure}

First, we examine how the number of additional tasks affects the quality of the multiple-$f_0$ output.
For this experiment, we compare the single-task multiple-$f_0$ model, a multitask model trained with two tasks (multiple-$f_0$ and vocal-$f_0$), a multitask model trained with three tasks (multiple-$f_0$, vocal-$f_0$, and bass), and finally the full four-task multitask model.
The results of this experiment are shown in Figure~\ref{fig:mtask-combos}, and we see that the multiple-$f_0$ results improve slightly with each additional task.
Note that we do not look at the change in performance for vocal-$f_0$, bass or melody because when the tasks is removed, we cannot measure the output.

\begin{figure}
\begin{center}
\includegraphics[width=\columnwidth]{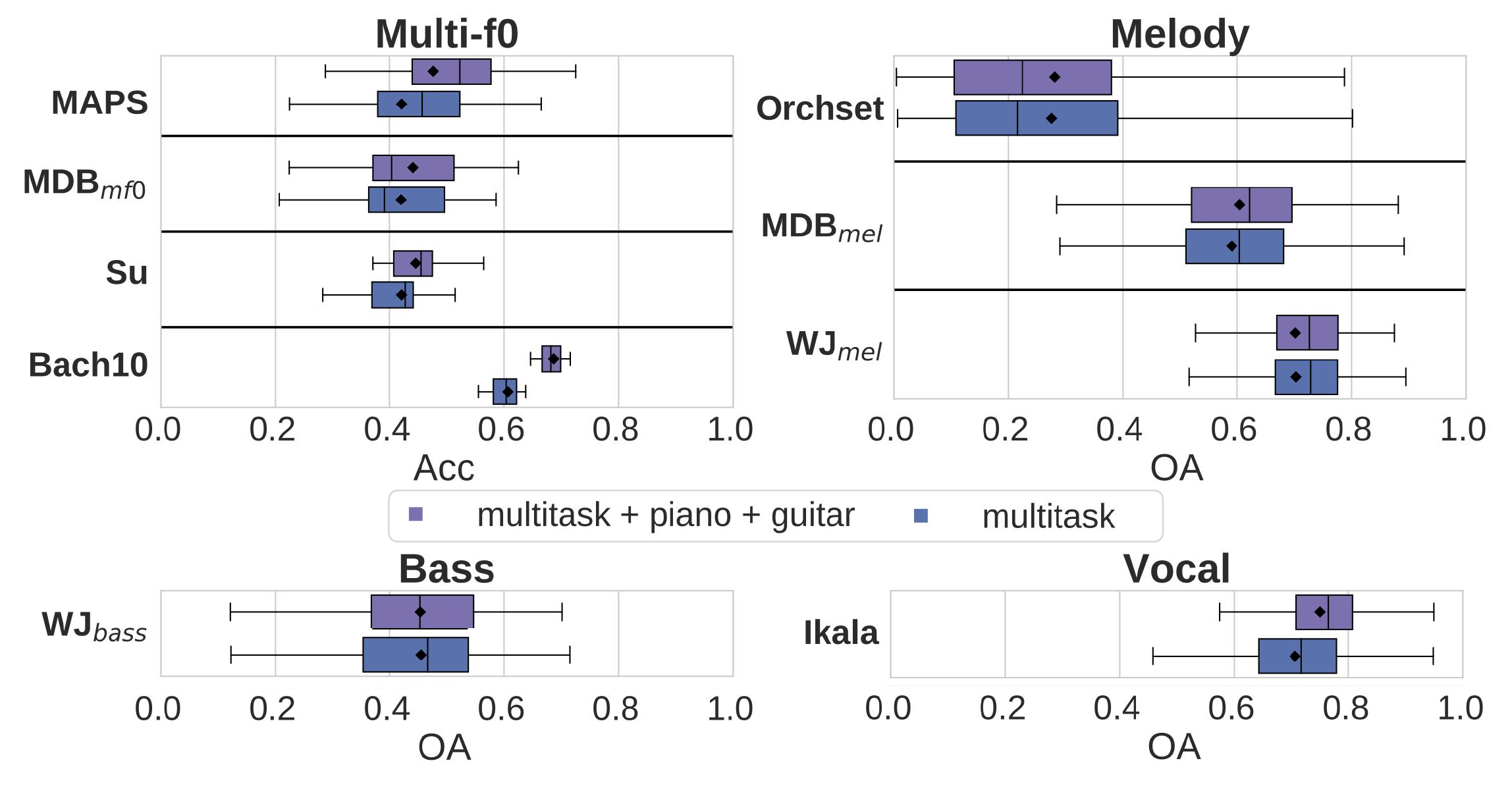}
\end{center}
\caption{The effect of adding additional tasks (piano and guitar), on the performance of each of the four main tasks.}\label{fig:mtask-artificial}
\end{figure}

To further explore this trend, we train a six-task model, where the additional two tasks are piano and guitar $f_0$ estimation, using the same set of audio training data as the 4-task model.
Note that the only audio data we have labeled for piano and guitar contains synthesized pianos and guitars that are mixed with ``real'' stems as described in Section~\ref{sec:artifical-mix}.
Figure~\ref{fig:mtask-artificial} compares these two models.
Overall, we see that once again, having more tasks improves performance for most datasets.
When it does not improve, the difference between the two models is negligible.
The improvement is most apparent for multiple-$f_0$, likely because piano and guitar often play less salient notes, and this forces the model to emphasize this content.

\begin{figure}
\begin{center}
\includegraphics[width=\columnwidth]{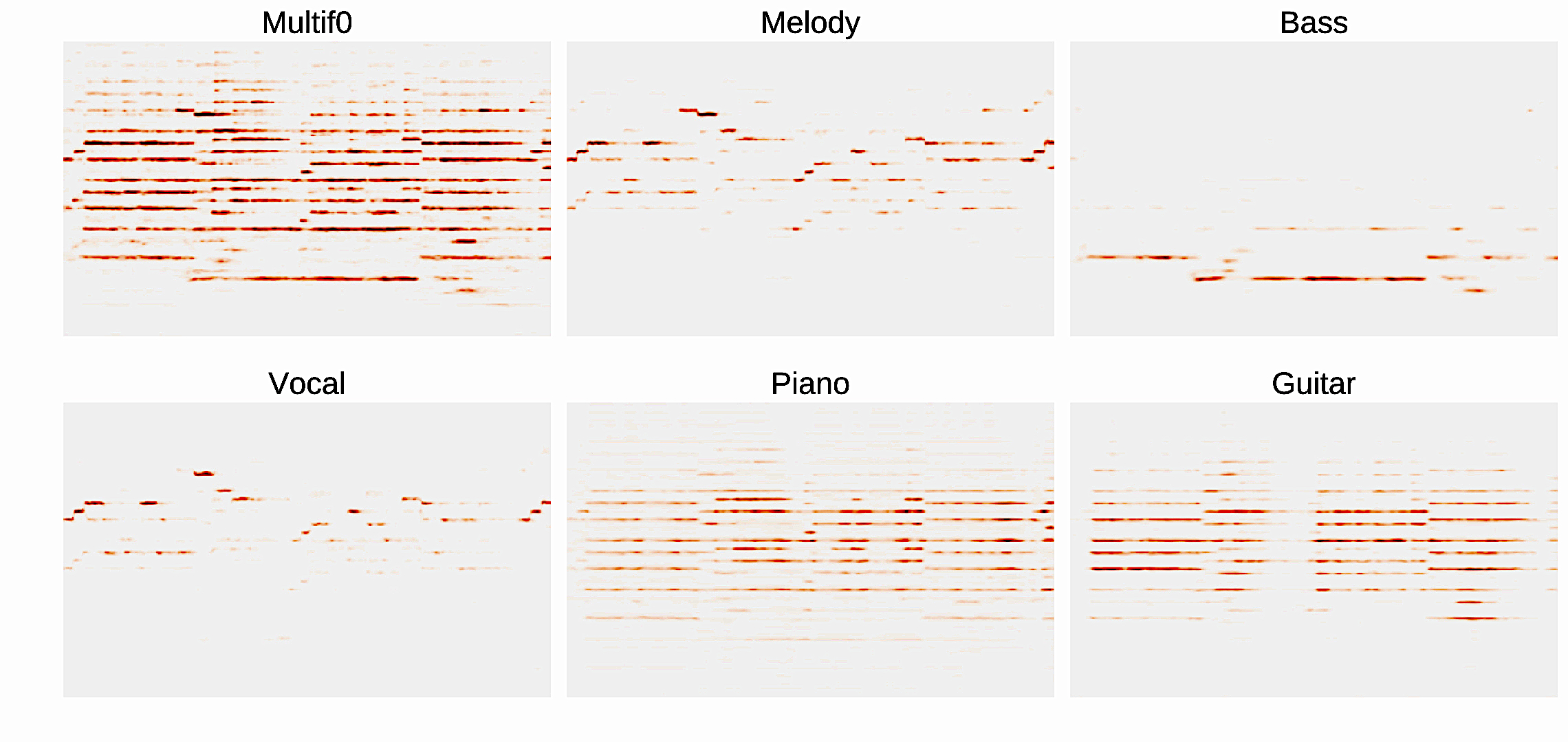}
\end{center}
\caption{Example of output produced by the 6-task multitask model. The excerpt shown is 10 seconds from an orchestral recording of the 3rd movement of Haydn's Symphony No. 94 (\textbf{Orchset}). Note that there are no vocals, piano, or guitar in this excerpt. As in Figure~\ref{fig:mtask-v_stask-qual}, the y-axis is plotted on a log-scale from C1 to C6.}\label{fig:mtask-artificial-qual}
\end{figure}

Figure~\ref{fig:mtask-artificial-qual} shows the predictions from the 6-task multitask model on an excerpt from \textbf{Orchset}.
The excerpt is of an orchestra and contains no vocals, piano, or guitar.
Interestingly, while the input audio does not contain these sources, the output produced for vocal, piano, and guitar look like what a voice, piano and guitar might produce.
For example, the guitar output has many note onsets occurring more or less the same time (e.g. like a strum), while the piano has a combination of block harmonies and moving lines.
This again suggests that the models are not good at distinguishing $f_0$ content based on timbre/harmonic patterns, but are instead modeling patterns of $f_0$ information.

\subsubsection{The Role of Synthesized Training Data}\label{sec:synth-data}

\begin{figure}
\begin{center}
\includegraphics[width=\columnwidth]{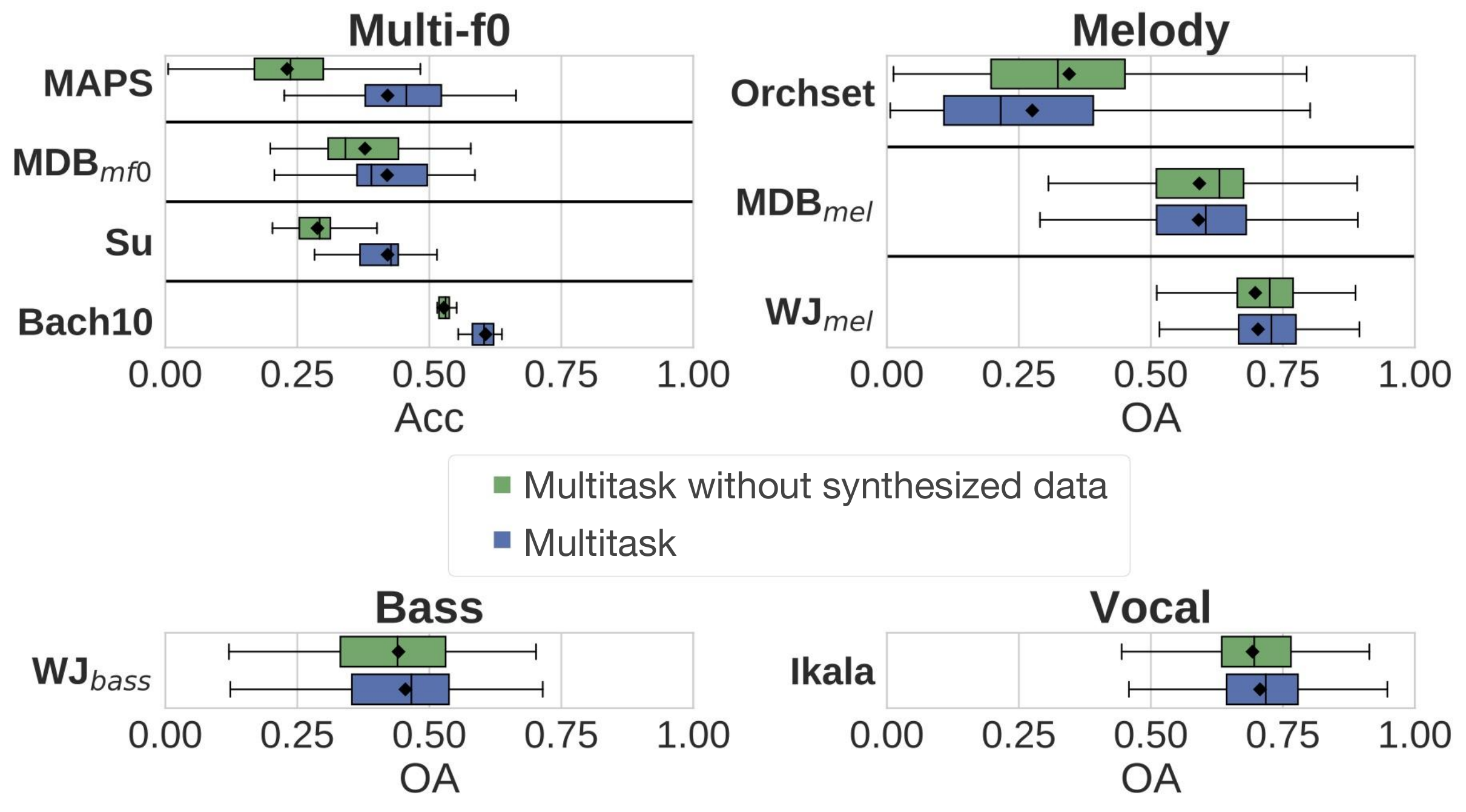}
\end{center}
\caption{Multitask model trained with and without synthesized training data (\texttt{RMX-P} and \texttt{RMX-PG}).}\label{fig:synth-dat}
\end{figure}

In this experiment, we explore the effect of adding synthesized piano and guitar stems in the training data (\texttt{RMX-P} and \texttt{RMX-PG}).
To do this, we train a multitask model on only \texttt{ORIG} and \texttt{RMX} data, and compare it with the multitask model trained on all data.
As shown in Figure~\ref{fig:synth-dat}, the synthesized data is beneficial for multiple-$f_0$ performance.
As discussed in Section~\ref{sec:training-data}, the audio and corresponding multiple-$f_0$ annotations for \texttt{RMX} are sparse (often only vocals, bass and drums), so it is unsurprising that training using data with a similar density to the test sets improves performance.

For melody, bass and vocal-$f_0$, the addition of synthesized training data does not have a significant effect, with the exception of \textbf{Orchset} where the addition of synthesized data hurts.
\textbf{Orchset} is the only dataset in the melody, bass and vocal test sets that does not contain any piano or guitar, thus the addition of the synthesized data is changing the distribution of the training set away from the distribution of the test set.

This result is particularly exciting - suggesting that training on synthesized music can generalize to real-world recordings.
However, we never trained on purely synthesized audio.
We suspect that this is why our model does not overfit the synthesized sound fonts, and expect that training on music where every source was synthesized would result in overfitting. 

\subsubsection{The Harmonic CQT}\label{sec:exp-hcqt}
In this experiment we examine the effectiveness of the Harmonic CQT as an input representation -- specifically we ask if the layered harmonics improve the performance of our model.
To do this we train four additional multitask models each with fewer (4, 3, 2, and 1) harmonics in the input representation.

\begin{figure}
\begin{center}
\includegraphics[width=\columnwidth]{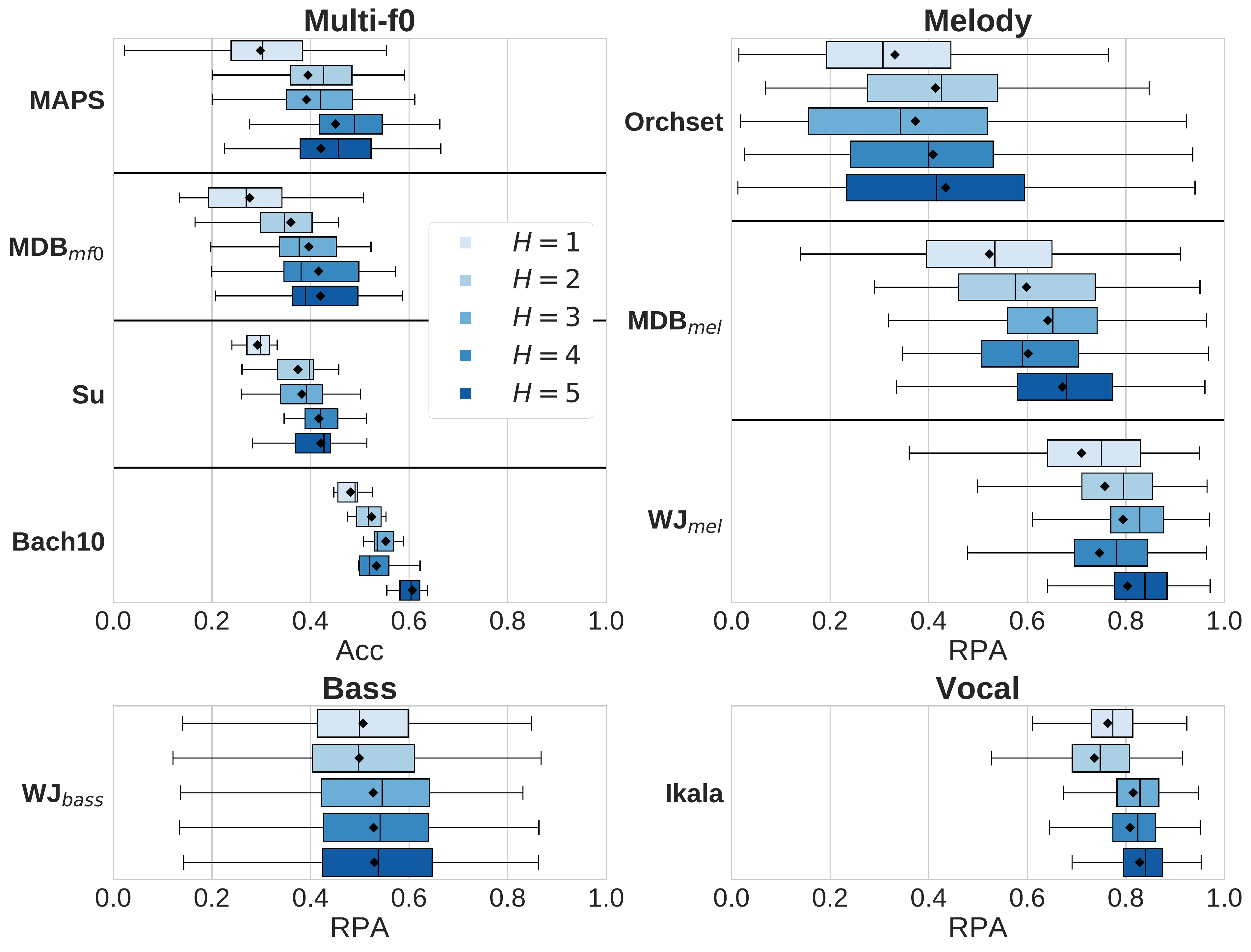}
\end{center}
\caption{The impact of removing harmonics in the HCQT on each task's performance, trained using the four-task multitask architecture.}\label{fig:nharms}
\end{figure}

Figure~\ref{fig:nharms} shows performance across the number of harmonics.\footnote{Note that for the single-$f_0$ tasks we are reporting Raw Pitch Accuracy (\texttt{RPA}) rather than Overall Accuracy because there is a stronger effect.
However, the trend is the same for Overall Accuracy.}
With the exception of bass, adding more harmonics to the input representation improves performance overall.
We expect that bass performance does not substantially improve with additional harmonics because most bass signals concentrate energy at the first harmonic, and the other weaker harmonics are in the range of other instruments, which become hard to detect in polyphony.

\subsection{Baseline Comparisons}\label{sec:exp-baselines}
In these experiments we compare the multiple-$f_0$, melody, bass and vocal $f_0$ outputs produced by our multitask architecture against strong baseline algorithms.

\begin{figure}
\begin{center}
\includegraphics[width=\columnwidth]{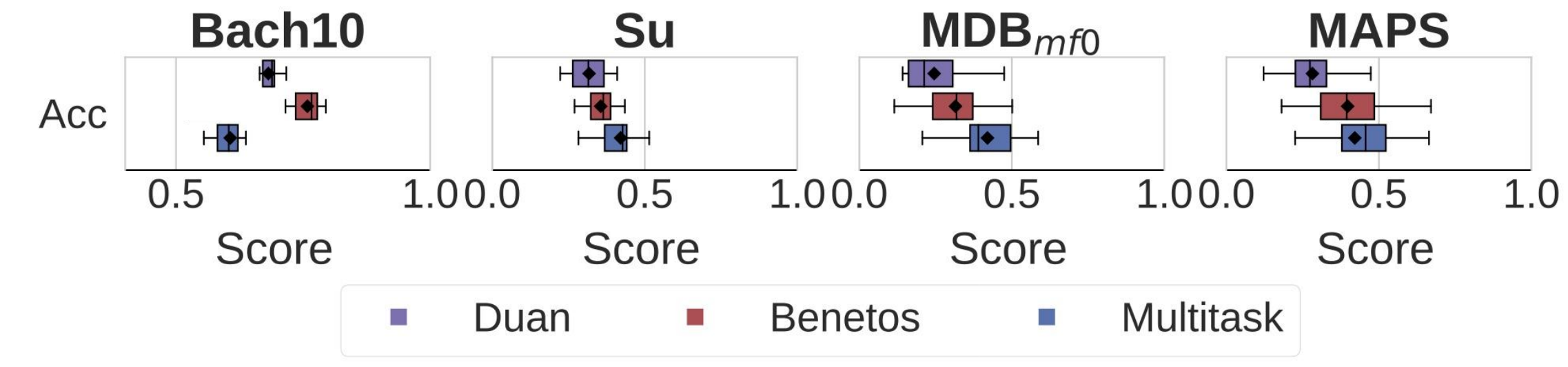}
\end{center}
\caption{Multitask multiple-$f_0$ output (blue) compared with baseline algorithm outputs by Duan~\cite{duan2010multiple} (purple) and Benetos~\cite{benetos2015efficient} (red).
In the upper plots, higher values for the metrics indicate better performance, while in the lower plots, lower values indicate better performance.}\label{fig:mf0-v-sota}
\end{figure}

First we compare the multitask multiple-$f_0$ output against the methods by Benetos~\cite{benetos2015efficient} and Duan~\cite{duan2010multiple} in Figure~\ref{fig:mf0-v-sota}.
Similar to the results of~\cite{bittner2017deep}, the multitask model outperforms the baselines on \textbf{MAPS}, \textbf{Su} and \textbf{MDB}$_{mf0}$, but under-perform on \textbf{Bach10} because the majority of the $f_0$ values produced by clarinet are not detected.

\begin{figure}
\begin{center}
\includegraphics[width=\columnwidth]{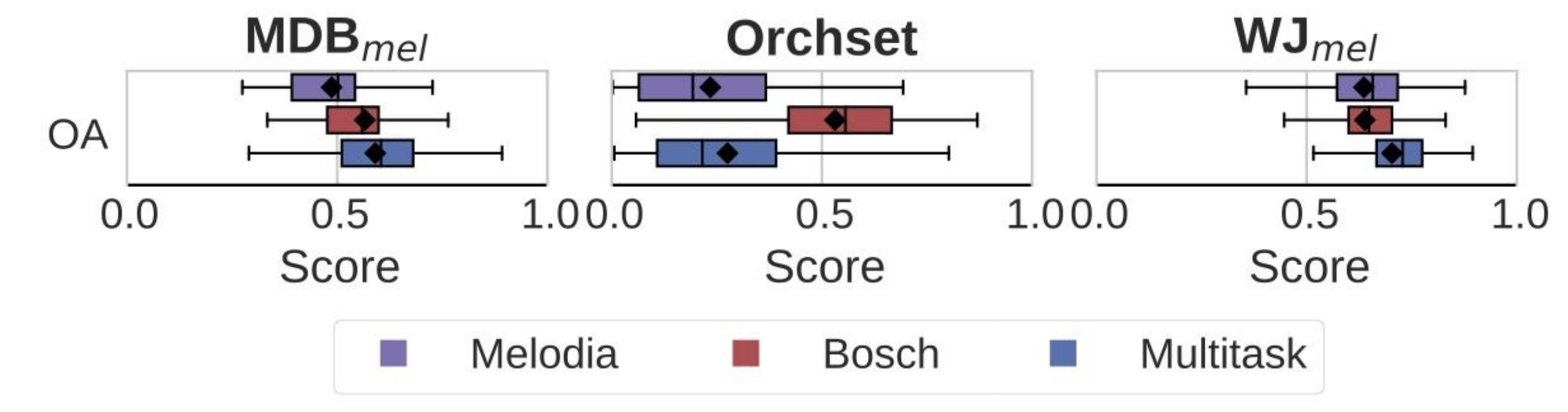}
\end{center}
\caption{Multitask melody output (blue) compared with baseline algorithm outputs by Salamon~\cite{Salamon:2014fh} (purple) and Bosch~\cite{bosch2017phd} (red).}\label{fig:mel-v-sota}
\end{figure}

Next, we compare the multitask melody output against two strong baselines by Salamon (``Melodia'')~\cite{salamonMelodyJournal} and a more recent algorithm by Bosch~\cite{bosch2017phd}, shown in Figure~\ref{fig:mel-v-sota}.
On \textbf{MDB}$_{mel}$ and \textbf{WJ}$_{mel}$ the multitask model outperforms the others, but not on \textbf{Orchset}.
\textbf{Orchset} is the test set that is most different from the training data, so it is not surprising that our model does not perform as well in that musical context.
Additionally, Bosch's model was designed specifically to capture the specificities in orchestral music, while ours was trained in a more generic musical context.

\begin{figure}
\begin{center}
\includegraphics[width=\columnwidth]{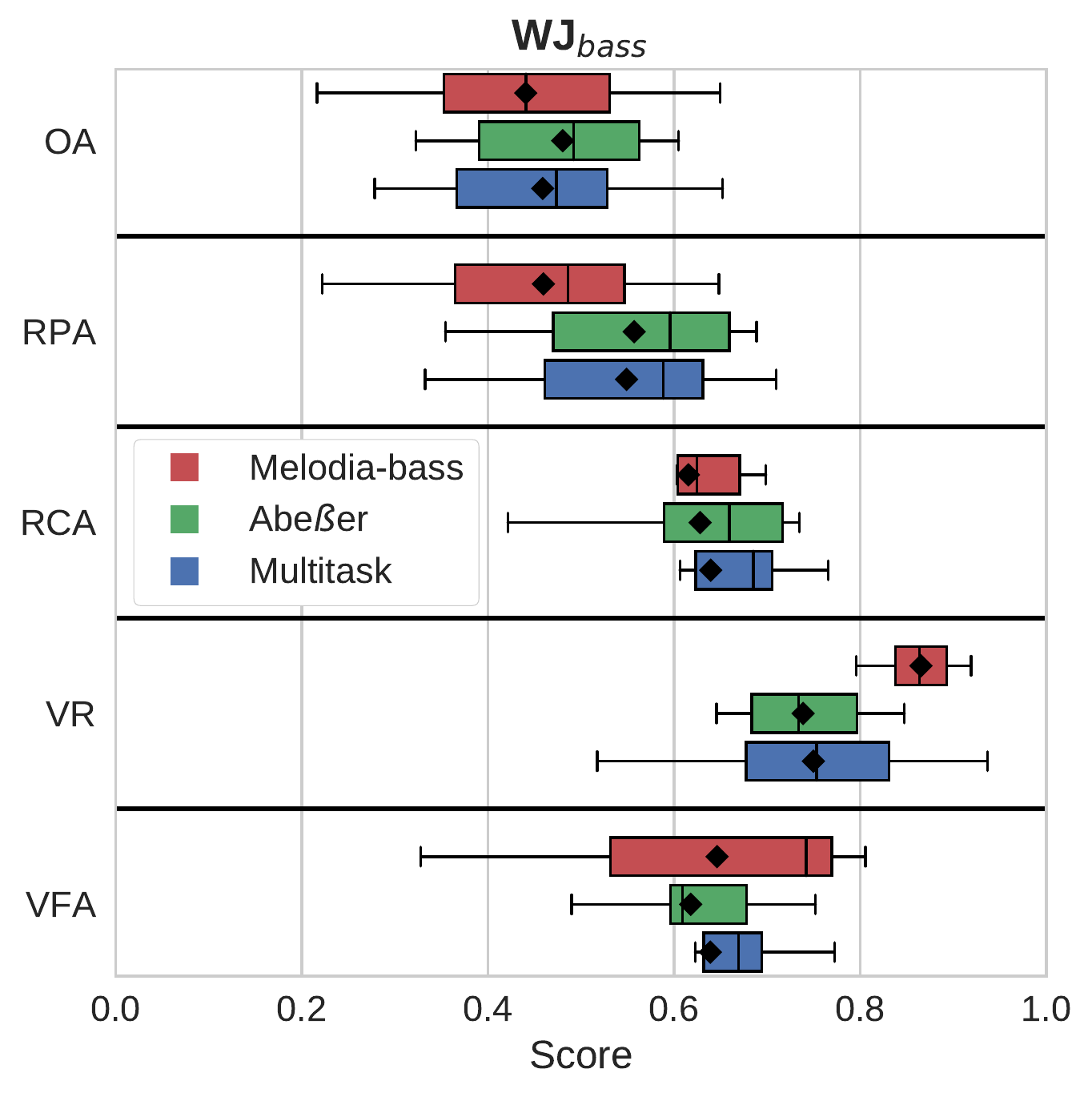}
\end{center}
\caption{Multitask bass output (blue) compared with baseline algorithms Melodia-bass~\cite{salamon2013tonal} (red) and Abe{\ss}er~\cite{abesser2017bass} (green) on a 10-track subset of \textbf{WJ}$_{bass}$. The other 30 tracks of \textbf{WJ}$_{bass}$ were excluded from this evaluation because they were used as training data for the algorithm by Abe{\ss}er.
}\label{fig:bass-v-sota}
\end{figure}

For bass, we compare the multitask output against two baseline algorithms by Salamon (Melodia-bass)~\cite{salamon2013tonal} and a similar salience-learning approach by Abe{\ss}er~\cite{abesser2017bass} on a 10-track subset of \textbf{WJ}$_{bass}$, shown in Figure~\ref{fig:bass-v-sota}.
We produce $f_0$ output from Abe{\ss}er's predicted salience representations in the same manner as for our salience representations: we select the frequency bin with maximum salience for each time frame and set frames that fall below a threshold (selected from the validation set) as unvoiced.
Note that we exclude the other 30 \textbf{WJ}$_{bass}$ tracks from this evaluation because Abe{\ss}er's salience representations are trained on this data.
None of the results are substantially different from each other, but on average Abe{\ss}er's algorithm outperforms the others.
This is not surprising as the training data is quite similar to this test set.

\begin{figure}
\begin{center}
\includegraphics[width=\columnwidth]{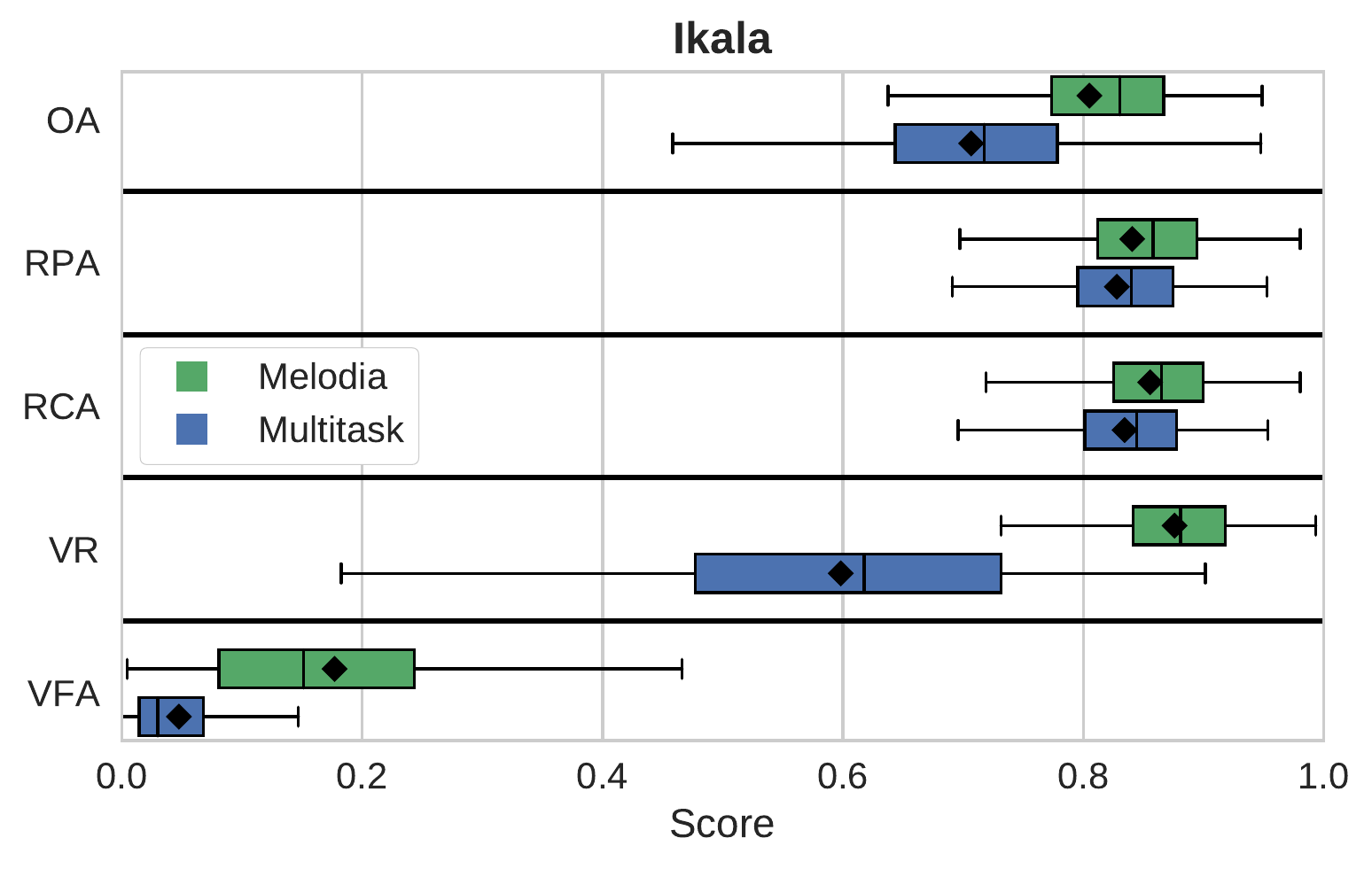}
\end{center}
\caption{Multitask vocal-$f_0$ output (blue) compared with Melodia~\cite{salamonMelodyJournal} (green) on \textbf{Ikala}.}\label{fig:voc-v-sota}
\end{figure}

Finally, for vocals we compare the multitask vocal output with Melodia~\cite{salamonMelodyJournal}, which is primarily designed for vocal data.
The results are shown in Figure~\ref{fig:voc-v-sota}, and we see that Melodia performs very well on \textbf{Ikala}, outperforming the multitask model by a significant margin in \texttt{OA}.
This overall difference is almost entirely a function of voicing: the multitask model has much lower \texttt{VR} than melodia, meaning that it misses lots of voiced frames; this again speaks to our finding that the multitask model fails to encode timbre information and thus is weak at determining accurate voicing information.
However, in terms of \texttt{RPA} and \texttt{RCA} the methods are much more comparable, with Melodia performing slightly better.

\section{Conclusions and Future Work}\label{sec:conclusions}
In this article, we presented a multitask model that jointly performs different kinds of $f_0$ estimation, and exploits the superset relationship between multiple-$f_0$ estimation and other $f_0$ tasks such as melody estimation.
We showed that overall, the multitask model outperforms its single task equivalents for multiple tasks across multiple datasets.
Next, we saw that the more tasks included in the multitask model, the higher the performance, and that this was even true for tasks trained with purely synthesized data.
Finally, we found that the addition of synthesized piano and guitar audio in the training set was beneficial for multiple-$f_0$ results and did not have a substantial effect on the other tasks.

A weakness observed in this model is that timbre information is not properly characterized, and thus the model focuses on the topology of $f_0$ information, resulting in confusions in particular between non-vocal melodic sources and vocals.
In the proposed model, the only timbre information available is learned from the multiple-$f_0$-masked HCQT.
While this allows the model to focus on harmonic series information, it masks all information outside of the local harmonic region, such as wide band noise that is often present during note onsets.
A potential way to mitigate this problem is to additionally predict pitch-localized onsets using features that are not masked.
Instrument identification could also be added as an additional task after the multiple-$f_0$ masking, forcing the latent feature representation to be separable along instrument classes.

Overall, the bass-$f_0$ estimation is weak across iterations of the multitask models, and we strongly suspect this is a result of the training data - the ``ground truth'' bass annotations were nearly all estimated by a pitch tracker, and we observed that it made a non-negligible number of octave and voicing mistakes.
The voicing mistakes in particular are problematic during training, with both non-existent $f_0$ values being trained as active and real $f_0$ values being trained as inactive.
To improve this, future work could utilize automatic $f_0$ annotations as presented in~\cite{salamon2017automatic} to ensure a perfect correspondence between the audio and annotations.
Additionally the synthesized piano and guitar training data can benefit from improved transcription methods such as~\cite{hawthorne2018onsets}, which predicts pitch-specific onsets along with continuous $f_0$ information.

\section*{Acknowledgements}
The authors would like to thank Juan Jos\'{e} Bosch, Justin Salamon, Jakob Abe{\ss}er, Emmanouil Benetos, and Zhiyao Duan for kindly making their algorithms available.

\bibliographystyle{spmpsci}
\bibliography{multitask}

%

\begin{IEEEbiography}[{\includegraphics[width=1in,clip,keepaspectratio]{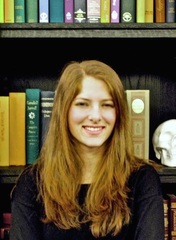}}]{Rachel M. Bittner}
Rachel is a Research Scientist at Spotify in New York City.
She completed her Ph.D. in May 2018 at the Music and Audio Research Lab at New York University under Dr. Juan P. Bello. Previously, she was a research assistant at NASA Ames Research Center working with Durand Begault in the Advanced Controls and Displays Laboratory. She did her master's degree in math at NYU's Courant Institute, and her bachelor's degree in music performance and math at UC Irvine. Her research interests are at the intersection of audio signal processing and machine learning, applied to musical audio. Her dissertation work applied machine learning to fundamental frequency estimation.
\end{IEEEbiography}

\begin{IEEEbiography}[{\includegraphics[width=1in,clip,keepaspectratio]{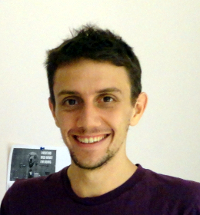}}]{Brian McFee}
Brian is an assistant professor in Music Technology and Data Science at New York University.
He develops machine learning tools to analyze multimedia data. This includes recommender systems, image and audio analysis, similarity learning, cross-modal feature integration, and automatic annotation.
From 2014 to 2018, he was a data science fellow at the Center for Data Science at New York University.
Previously, he was a postdoctoral research scholar in the Center for Jazz Studies and LabROSA at Columbia University.
Before that, he was advised by Prof. Gert Lanckriet in the Computer Audition Lab and Artificial Intelligence Group at the University of California, San Diego.
In May, 2012, he defended his dissertation, titled ``More like this: machine learning approaches to music similarity''.
\end{IEEEbiography}


\begin{IEEEbiography}[{\includegraphics[width=1in,clip,keepaspectratio]{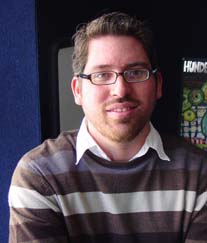}}]{Juan P. Bello}
Juan Pablo Bello is Associate Professor of Music Technology and Computer Science and Engineering at New York University. In 1998 he received a BEng in Electronics from the Universidad Sim{\'o}n Bolívar in Caracas, Venezuela, and in 2003 he earned a doctorate in Electronic Engineering at Queen Mary, University of London. Juan’s expertise is in digital signal processing, machine listening and music information retrieval, topics that he teaches and in which he has published more than 100 papers and articles in books, journals and conference proceedings. He is director of the Music and Audio Research Lab (MARL), where he leads research on music informatics. His work has been supported by public and private institutions in Venezuela, the UK, and the US, including Frontier and CAREER awards from the National Science Foundation and a Fulbright scholar grant for multidisciplinary studies in France.
\end{IEEEbiography}




\end{document}